%% file: main.tex
\documentclass[10pt,twocolumn,letterpaper]{article}

\usepackage{cvpr}              % To produce the CAMERA-READY version
\input{preamble}
\definecolor{cvprblue}{rgb}{0.21,0.49,0.74}
\usepackage[pagebackref,breaklinks,colorlinks,allcolors=cvprblue]{hyperref}
\usepackage{amsmath,amssymb,amsthm,mathtools}
\usepackage{bm}
\usepackage{xcolor}
\usepackage{enumitem}
\usepackage{booktabs}
\usepackage{multirow}

\def\confName{CVPR}
\def\confYear{2026}

\title{IrisFP: Adversarial-Example-based Model Fingerprinting with Enhanced Uniqueness and Robustness}

\author{
Ziye Geng\textsuperscript{\rm 1},
Guang Yang\textsuperscript{\rm 2},
Yihang Chen\textsuperscript{\rm 1},
and Changqing Luo\textsuperscript{\rm 1},\\
\textsuperscript{\rm 1}University of Houston,
\textsuperscript{\rm 2}Virginia Commonwealth University  \\
{\tt\small \{zgeng2, ychen165, cluo3\}@uh.edu,}
{\tt\small yangg2@vcu.edu}
}

\begin{document}
\maketitle
\input{sec/0_abstract}    
\input{sec/1_intro}
\input{sec/2_preliminary}

\input{sec/3_methodology}
\input{sec/4_experiments}

\input{sec/5_conclusion}

{
    \small
    \bibliographystyle{ieeenat_fullname}
    \bibliography{main}
}

  \input{sec/6_appendix}

\end{document}

%% file: sec/0_abstract.tex
\begin{abstract}

We propose IrisFP, a novel adversarial-example-based model fingerprinting framework that enhances both uniqueness and robustness by leveraging multi-boundary characteristics, multi-sample behaviors, and fingerprint discriminative power assessment to generate composite-sample fingerprints. Three key innovations make IrisFP outstanding: 1) It positions fingerprints near the intersection of all decision boundaries—unlike prior methods that target a single boundary—thus increasing the prediction margin without placing fingerprints deep inside target class regions, enhancing both robustness and uniqueness; 2) It constructs composite-sample fingerprints, each comprising multiple samples close to the multi-boundary intersection, to exploit collective behavior patterns and further boost uniqueness; and 3) It assesses the discriminative power of generated fingerprints using statistical separability metrics developed based on two reference model sets, respectively, for pirated and independently-trained models, retains the fingerprints with high discriminative power, and assigns fingerprint-specific thresholds to such retained fingerprints. Extensive experiments show that IrisFP consistently outperforms state-of-the-art methods, achieving reliable ownership verification by enhancing both robustness and uniqueness.

\end{abstract}

%% file: sec/1_intro.tex
%%%%%%%%%%%%%%%%%%%%%%%%%%%%%%%%%%%%
\section{Introduction}
\label{sec:requirement}
%%%%%%%%%%%%%%%%%%%%%%%%%%%%%%%%%%%%

Adversarial-example-based model fingerprinting has recently gained attention as a promising approach to protect intellectual property (IP) of deep neural networks (DNNs) developed for many tasks like image classification. This type of model fingerprinting builds on the adversarial example technique to generate input-output pairs as fingerprints that exhibit distinctive model-specific behavior. More specifically, such a fingerprinting approach introduces subtle perturbations to clean inputs to elicit unique model-specific responses from a protected model: the protected model produces outputs that deviate from those induced by the original clean inputs, whereas independently-trained models tend to produce outputs consistent with the clean inputs and significantly different from those of the protected model \cite{IPIP21, DNNF21, FDNN22, AAFA20, FITP25,RFRF2025}. These crafted fingerprints are then utilized for ownership verification~\cite{TMFA21,DMFA24}.

So far, researchers have developed many adversarial-example-based model fingerprinting methods \cite{IPIP21,FDNN22,UWSDW24,QURD25,lin2025adversarial,MCGF24,yang2026anafp,yang2026liteguard}, which typically generate individual fingerprints near a single decision boundary, without considering their relative distances to other boundaries.
For example, IPGuard \cite{IPIP21} crafts fingerprints close to the decision boundaries of the protected model to capture model-specific behaviors; UAP \cite{FDNN22} leverages universal perturbations to induce model-dependent outputs over a broad set of inputs; ADV-TRA \cite{UWSDW24} constructs adversarial trajectories composed of samples that traverse boundary regions to capture rich model characteristics; and AKH \cite{QURD25} constructs fingerprints by generating adversarial examples that originate from naturally misclassified samples. 

\begin{figure}[htbp]
    \centering    \includegraphics[width=0.71\linewidth]{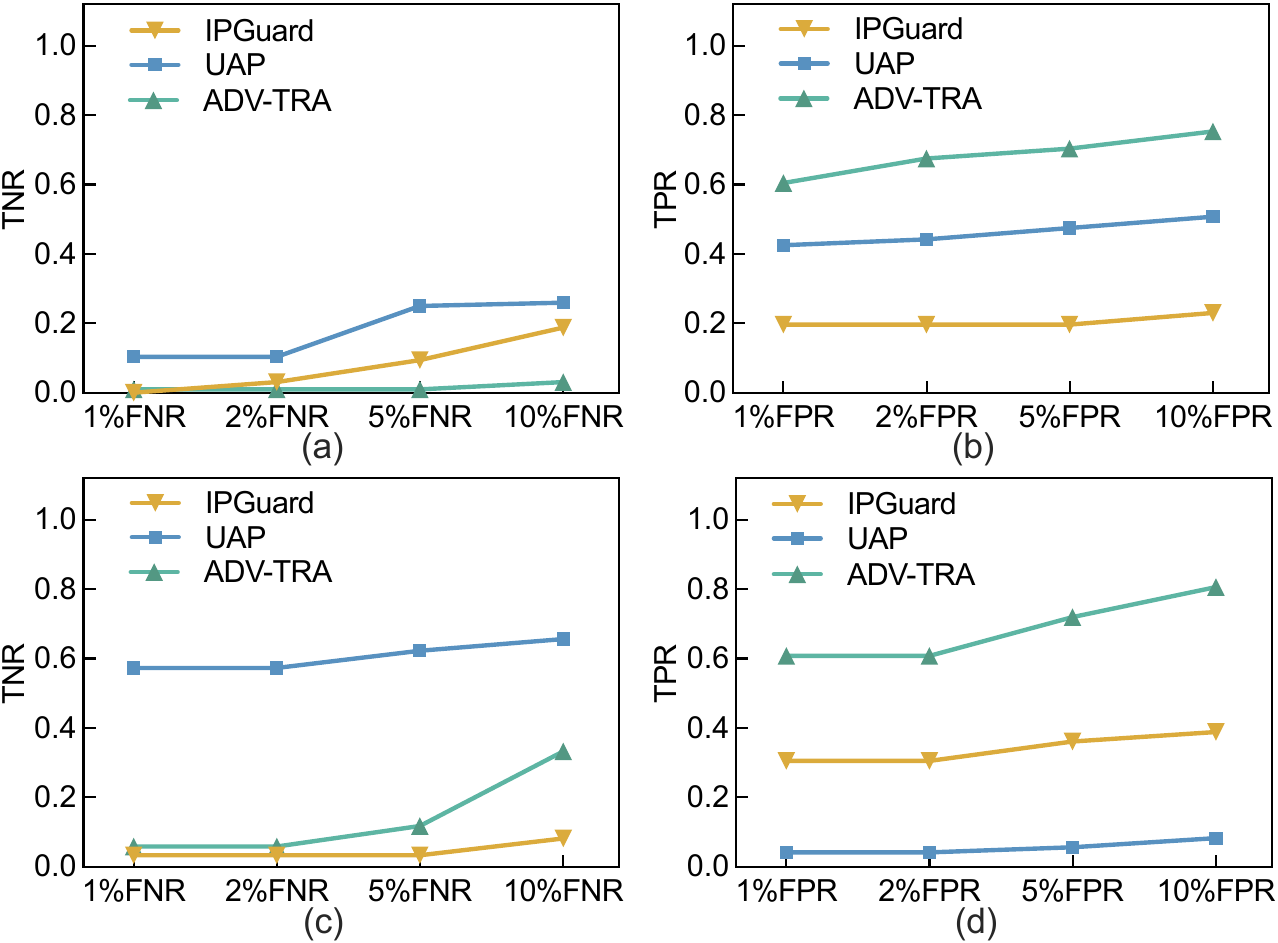}
  %  \vspace{-0.1in}
    \caption{The uniqueness and robustness achieved by prior adversarial-example-based fingerprinting methods.}
\label{fig:robust_unique_motivation}
\end{figure}

However, adversarial-example-based model fingerprinting is inherently prone to model modification attacks like fine-tuning and pruning, as such attacks can shift the decision boundaries of protected models, consequently circumventing ownership verification \cite{CRAT21,HAAAH20, MEAR24}. To address this issue, prior methods propose placing fingerprints deep inside the regions of target classes—far away from decision boundaries—to enhance robustness against model modifications \cite{IPIP21,MCGF24}. Nevertheless, this strategy may in turn undermine uniqueness, as such fingerprints are less sensitive to model-specific decision boundaries. Consequently, existing methods have weak uniqueness or robustness, or even both. This observation has been corroborated by our experimental results in Figure~\ref{fig:robust_unique_motivation}. The experiments, performed on CIFAR-10 and Fashion-MNIST, assess uniqueness by measuring true negative rates (TNRs) under varying false negative rate (FNR) levels, and robustness by measuring true positive rates (TPRs) under different false positive rate (FPR) levels. A higher TNR indicates better uniqueness in distinguishing the protected model from independently-trained ones, while a higher TPR indicates stronger robustness to model modifications. The results show that conventional methods can achieve either weak uniqueness or robustness, which highlights the inherent difficulty of generating effective fingerprints for ownership verification.

%%%%%%%%%%%%%%%%%%%%%%%%%%%%%%%%%
\noindent \textbf{Main question}. Based on the discussion so far, we have the following research question: \textit{Is it possible to design a new adversarial-example-based model fingerprinting method that enhances both uniqueness and robustness?}

%%%%%%%%%%%%%%%%%%%%%%%%%%%%%
\subsection{Our Answer}
% \subsection{Our results}
\label{subsec:result}
%%%%%%%%%%%%%%%%%%%%%%%%%%%%

In this paper, we give an affirmative answer to the above question. Recent studies \cite{IBSF24,SDBF25} show that compared to fingerprints positioned on a single decision boundary, those placed at the intersection of multi-class decision boundaries has high sensitivity, i.e., achieving enhanced uniqueness. By leveraging this property, IBSF \cite{IBSF24} and SDBF \cite{SDBF25} were developed to detect model tampering. While the fingerprints placed at the intersection of multi-class decision boundaries can enhance uniqueness, they are extremely vulnerable to model modifications, i.e., struggling with weak robustness. Therefore, our question is: \textit{Can we exploit decision boundary intersection to design a new fingerprinting method with enhanced uniqueness and robustness?}

\begin{figure}[htbp]
    \centering
    \includegraphics[width=0.6\linewidth]{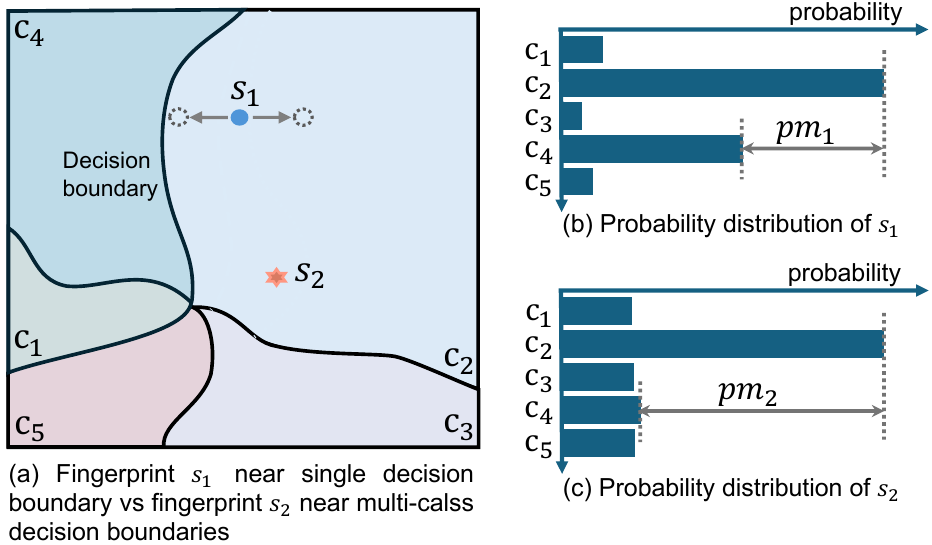}
   % \vspace{-0.1in}
    \caption{(a): The placement of input samples $s_1$ and $s_2$ in a region; (b): The prediction margin $pm_1$ of fingerprint $s_1$; and (c): The prediction margin $pm_2$ of fingerprint $s_2$.}
    \label{fig_problem}
\end{figure}

To positively answer this question, we first provide an intuitive illustration by Figure~\ref{fig_problem} to show that proximity to all decision boundaries is crucial for enhancing robustness. As illustrated in Figure~\ref{fig_problem}(a), sample $s_1$ lies within the region of class $c_2$, positioned near the boundary between $c_2$ and $c_4$, but is far from the other regions, while $s_2$, which also belongs to $c_2$, is located near the intersection of multiple boundaries, making it simultaneously close to all surrounding class regions. The corresponding output distributions for $s_1$ and $s_2$, shown in Figure~\ref{fig_problem}(b) and Figure~\ref{fig_problem}(c), reveal that $s_2$ has a larger prediction margin—that is, a greater confidence gap between the top two predicted classes. This increased margin means enhancing robustness at no cost of uniqueness. This key observation motivates our design: to construct fingerprints that are not only model-specific but also strategically placed near the multi-boundary intersection to improve uniqueness and robustness.

\textbf{Our Design:} We propose IrisFP, a novel model fingerprinting framework that generates un\underline{I}que and \underline{r}obust compos\underline{i}te-\underline{s}ample \underline{F}inger\underline{P}rints (IrisFP) by exploiting multi-boundary characteristics, multi-sample behaviors, and fingerprint quality assessment. Unlike conventional methods that place each fingerprint near a single decision boundary, IrisFP first generates fingerprint seeds by crafting adversarial samples to reside near the intersection of all decision boundaries of a protected model. This strategy increases the prediction margin without placing the sample deep inside the target class region, thus enhancing both robustness and uniqueness. To further boost uniqueness, IrisFP applies subtle, diverse perturbations to each fingerprint seed, generating multiple variants which, together with the seed, collectively form a composite-sample fingerprint.
Compared to the original sample, the seed and its variants are crafted to elicit different responses from the protected model, while tending to produce consistent predictions by independently-trained ones, resulting in a clear behavioral gap that enhances the fingerprint's discriminative capability. 
Moreover, IrisFP addresses a key limitation in prior adversarial-example-based fingerprinting methods—their lack of consideration for model modifications and independently trained models during fingerprint construction. To overcome this, IrisFP employs a fingerprint refinement strategy that evaluates the discriminative power of each fingerprint using statistical separability metrics to retain the fingerprints with high discriminative power, and assigns sample-specific thresholds to the retained fingerprints. 
Finally, IrisFP performs model ownership verification through a two-step process involving ownership matching and decision aggregation, specifically designed for composite-sample fingerprints.

\begin{figure*}[htbp]
    \centering
    \includegraphics[width=0.87\textwidth]{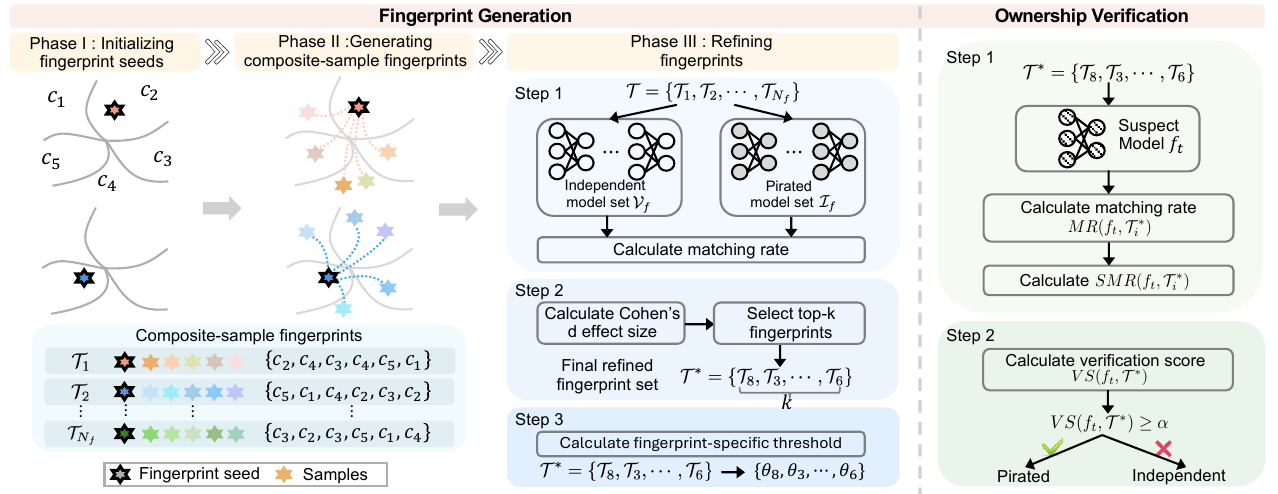}
   % \vspace{-0.1in}
    \caption{The overview of IrisFP.}
    \label{fig:workflow}
\end{figure*}

%% file: sec/2_preliminary.tex
\section{Preliminary}
\label{sec:preliminary}
%%%%%%%%%%%%%%%%%%%%%%%%%%%%%%%%%%%%%%%%%%
%%%%%%%%%%%%%%%%%%%%%%%%%%%%%%%%%%%
\subsection{Model Fingerprinting}
%%%%%%%%%%%%%%%%%%%%%%%%%%%%%%%%%%%

An adversarial-example-based fingerprinting approach typically allows a model owner to craft inputs that elicit unique model-specific responses. Given a protected model $f$ and a clean input $x_i$ with ground-truth label $y_i$, a small perturbation $\delta_i$, which is constrained to capture model-specific boundary characteristics, is added to produce a perturbed input $\hat{x}_i = x_i + \delta_{i}$, such that $f$ outputs a target label $\hat{y}_i \neq y_i$, resulting in fingerprint $(\hat{x}_i, \hat{y}_i)$. We mathematically formulate this process by
\begin{equation}
\min \|x_i - \hat{x}_i\|, \quad \text{s.t.} \quad f(\hat{x}_i) \neq {y_i}, %\ne y_i,
\label{eq:adv_example}
\end{equation}
where $\|\cdot\|$ represents a distance metric (e.g., $\ell_2$ or $\ell_\infty$ norm). This process can generate a set of fingerprints whose inputs are typically located near the decision boundaries of model $f$, making the fingerprints' behavior highly model-specific. The resulting fingerprints are used to query a target model $f_t$ to produce query results that are compared to the expected outputs for determining ownership.

%%%%%%%%%%%%%%%%%%%%%%%%%%%%%%%%%%%%%
\subsection{Model Modification Attacks}
%%%%%%%%%%%%%%%%%%%%%%%%%%%%%%%%%%%%%

In real-world scenarios, an adversary may alter a protected model $f$ through model modification techniques to create a pirated variant $f_p$ in order to evade ownership verification. Common modification techniques include fine-tuning (FT) \cite{CSTL20, TFVVI23}, pruning (PR) \cite{DCCDN15, PFEC17}, adversarial training (AT) \cite{BMRVI24, ATFF19}, and knowledge distillation (KD) \cite{OFTA24}. In practice, adversaries may also combine these techniques—such as injecting noise or pruning before fine-tuning—to amplify the degree of modification. Such alterations can shift the model's decision boundaries, potentially invalidating existing fingerprints and thereby enabling evasion of verification \cite{IPAGM24}. For example, given a fingerprint $(\hat{x}_i, \hat{y}_i)$ for the protected model $f$, querying $f_p$ with $\hat{x}_i$ might yield an output differing from the expected one, i.e., $f_{p}(\hat{x}_i) \neq \hat{y}_i = f(\hat{x}_i)$, thus undermining the reliability of ownership verification.

%% file: sec/3_methodology.tex
%%%%%%%%%%%%%%%%%%%%%%%%%%%%%%%%%%%%%%%%%%
\section{Methodology}
\label{sec:methodology}
%%%%%%%%%%%%%%%%%%%%%%%%%%%%%%%%%%%%%%%%%%

%%%%%%%%%%%%%%%%%%%%%%%%%%%
\subsection{Overview}
%%%%%%%%%%%%%%%%%%%%%%%%%%%

We propose IrisFP, a novel model fingerprinting framework that enhances both uniqueness and robustness. As shown in Figure~\ref{fig:workflow}, IrisFP comprises two main processes: fingerprint generation and ownership verification. The fingerprint generation process includes three phases: fingerprint seed initialization, composite-sample fingerprint generation, and fingerprint set refinement. The first phase places fingerprint seeds near the intersection of all decision boundaries to maximize prediction margins while keeping model-specific behavior intact. The second phase then minimally perturbs each seed to produce a set of variants with diverse outputs that, along with the seed, form a composite-sample fingerprint. The third phase further exploits statistical separability to select the most discriminative fingerprints from the generated ones to construct a fingerprint set, and assigns a specific threshold to each selected fingerprint.

The ownership verification process involves two steps: ownership matching and decision aggregation. At the first step, the matching rate across samples of each composite-sample fingerprint is computed and compared with its corresponding threshold to determine if the fingerprint is matched. At the second step, the matching outcomes across all fingerprints are aggregated to make a final decision. If the proportion of matched fingerprints exceeds a predefined threshold, the target model is deemed pirated; otherwise, it is identified as independently-trained.

%______________________________________
\subsection{The Fingerprint Generation Process}
%______________________________________
%
\subsubsection{Phase I: Initializing Fingerprint Seeds} Our proposed model fingerprinting framework starts with generating initial fingerprint seeds. Given a protected model $f$ trained on a dataset $\mathcal{D}$ with $C$ classes, we first randomly select $N_f$ input-label pairs $\{(x_i^0,y_i^0)\}_{i=1}^{N_f}$ from $\mathcal{D}$, where $x_i^0$ is an input with ground-truth label $y_i^0$. For each $x_i^0$, our adversarial-example-based model fingerprinting method introduces a trainable perturbation $\delta_i^0$ to produce a fingerprint seed $\hat{x}_i^0 = x_i^0 + \delta_i^0$. Unlike traditional fingerprinting methods that directly push a fingerprint toward a single decision boundary associated with a specific target class, our method instead guides $\hat{x}_i^0$ to be located near the intersection of $f$'s decision boundaries. 
This design enhances robustness: positioning samples close to such an intersection enables $f$ to predict the target class $\hat{y}_i^0$ with high confidence, while distributing the remaining probability almost evenly across other classes. Thus, model-specific output behavior is preserved, and the prediction margin is increased, improving robustness to modifications without sacrificing uniqueness.

To obtain such seeds, we define a biased probability distribution $p_i \in \mathbb{R}^C$ for each sample with respect to a randomly chosen target class $\hat{y}_i^0$, with $p_i(y)$ denoting the probability assigned to class $y$:
\begin{equation}
p_i(y) = 
\begin{cases} 
\frac{1}{C} + \tau, & \text{if } y = \hat{y}_i^0, \\ 
\frac{1 - (\frac{1}{C} + \tau)}{C - 1}, & \text{otherwise},
\end{cases}
%\quad \text{where } \tau > 0
\label{eq:p1_distribution}
\end{equation}
where $0< \tau < 1$ is a tunable hyperparameter controlling the degree of bias toward $\hat{y}_i^0$. Then, we define $f_o(x)$ as the softmax output of the model $f$, which produces a probability distribution over $C$ classes. The fingerprint seed $\hat{x}_i^0$ is then optimized to ensure that the model's output distribution $f_o(\hat{x}_i^0)$ aligns with the predefined $p_i$. This is achieved by minimizing the Kullback-Leibler (KL) divergence between $f_o(\hat{x}_i^0)$ and $p_i$, with an $L_1$ regularization term applied to minimize the perturbation. The loss function is given by $\mathcal{L}_{\text{phase1}} = \text{KL}(f_o(\hat{x}_i^0) \, || \, p_i) + \lambda_1 \| \delta_i^0 \|_1$,
%\begin{equation}
    % \min_{\{\delta_i\}} \sum_{i=1}^{N_f} \text{KL}(f_o(\hat{x_i}) \, || \, p_i) + \lambda_1 \| \delta_i \|_1   
%    \label{eq:placeholder}
%\end{equation}
where $\lambda_1$ is a regularization coefficient.
This optimization process yields a set of fingerprint seeds $\{(\hat{x}_i^0, \hat{y}_i^0)\}_{i=1}^{N_f}$, each positioned near the intersection of multi-class decision boundaries.

%%%%%%%%%%%%%%%%%%%%%%%%%%%%%%%%%%%%%%%%%%%
\subsubsection{Phase II: Generating Composite-sample Fingerprints}
To enhance uniqueness, we extend each fingerprint seed into a composite-sample fingerprint comprising multiple minimally perturbed variants designed to elicit diverse responses from the protected model. %This process aims to construct a fine-grained behavioral signature that reflects the model’s unique decision boundaries.
Specifically, each fingerprint seed $\hat{x}_i^0$ is perturbed by a set of small, trainable perturbations $\{\delta_i^1, \delta_i^2, \dots, \delta_i^T\}$ to generate a set of variants $\{\hat{x}_i^1, \hat{x}_i^2, \cdots, \hat{x}_i^T\}$, where $\hat{x}_i^t = \hat{x}_i^0 + \delta_i^t$, for $1 \leq t \leq T$. Each variant is assigned a target class uniformly sampled from $\{1, 2, \cdots, C\}$.
To position each variant near the intersection of multiple decision boundaries, we define a biased target probability distribution $p_i^t$ for each variant toward a randomly chosen target class $\hat{y}_i^t$, and $p_i^t(y)$ is the probability assigned to class $y$:
\begin{equation}
p_i^t(y) = 
\begin{cases} 
\frac{1}{C} + \tau, & \text{if } y = \hat{y}_i^t, \\ 
\frac{1 - (\frac{1}{C} + \tau)}{C - 1}, & \text{otherwise}.
\end{cases}
%\quad \text{where } \tau > 0
\label{eq:p2_distribution}
\end{equation}
%where $0< \tau < 1$ is a hyperparameter controls the degree of bias toward $\hat{y}_i^t$.
%
The perturbations $\{\delta_i^1, \delta_i^2, \cdots, \delta_i^T\}$ are jointly optimized by minimizing the average KL divergence between the model’s output distribution and the biased targets. The loss function is formulated as $\mathcal{L}_{\text{phase2}} = \frac{1}{T} \sum_{t=1}^{T} \left[ \text{KL} \left(f_o (\hat{x}_i^t) \, \| \, p_i^t \right) + \lambda_2 \| \delta_i^t \|_1 \right]$,
where $\lambda_2$ is a regularization coefficient. This process yields a fingerprint set with $N_f$ composite-sample fingerprints $\{\mathcal{T}_i\}_{i=1}^{N_f}$, where $\mathcal{T}_i = \{(\hat{x}_i^0, \hat{y}_i^0), (\hat{x}_i^1, \hat{y}_i^1), \dots, (\hat{x}_i^T, \hat{y}_i^T)\}$. %, including the fingerprint seed $(\hat{x}_i^0, \hat{y}_i^0)$ and $T$ variants.

By applying minimal, model-specific perturbations to each seed, IrisFP captures subtle variations in the model's decision landscape. These perturbations are optimized to induce diverse responses, leveraging the fact that adversarial perturbations are highly sensitive to a model's decision boundaries. Since independently-trained models---even those with architectural similarity---often exhibit different decision boundaries, their outputs across all samples in a fingerprint are typically inconsistent with those of the protected model, failing to replicate the behavioral pattern elicited by the protected model. This behavioral discrepancy enhances the discriminative capability of composite-sample fingerprints. 
%______________________________________
\subsubsection{Phase III: Refining Fingerprints}
%______________________________________

Some of the generated composite-sample fingerprints may not have strong and consistent discriminative behaviors, inevitably degrading ownership verification performance. Therefore, we propose a fingerprint refinement strategy comprising three steps: calculating the matching rate, selecting composite-sample fingerprints, and computing fingerprint-specific thresholds.

Prior to describing these steps, we define two reference model sets used throughout the refinement process:
\begin{itemize}[left=0pt]
\item \textbf{Pirated Model Set} ($\mathcal{V}_f$): Models derived from the protected model via model modification techniques. %such as fine-tuning, pruning, adversarial training, or knowledge distillation.

\item \textbf{Independent Model Set} ($\mathcal{I}_f$): Models trained independently from scratch using different initializations, architectures, or training data.% and thus unrelated to the protected model.

\end{itemize}
Both sets are used to evaluate whether a composite-sample fingerprint elicits distinguishable responses between pirated and independently-trained models.

\textbf{Step 1: Calculating the matching rate.}
To support fingerprint refinement, we first compute the fingerprint-level matching rate $MR(f, \mathcal{T}_i)$ for each composite-sample fingerprint $\mathcal{T}_i$ across models in both $\mathcal{V}_f$ and $\mathcal{I}_f$. Given $\mathcal{T}_i = \{(\hat{x}_i^0, \hat{y}_i^0), (\hat{x}_i^1, \hat{y}_i^1), \dots, (\hat{x}_i^T, \hat{y}_i^T)\}$ and a model $f_j$ within $\mathcal{V}_f$ or $\mathcal{I}_f$, the fingerprint-level matching rate is defined as $\text{MR}(f_j, \mathcal{T}_i) = \frac{1}{T+1} \sum_{t=0}^{T}\mathbb{I}{[ f_j(\hat{x}_i^t) = \hat{y}_i^t]}$,
where $\mathbb{I}[\cdot]$ is an indicator function that returns ``1'' if $f_j$ with input $\hat{x}_i^t$ outputs $\hat{y}_i^t$, and ``0'' otherwise.
This metric quantifies the degree to which the model's behavior aligns with the protected model.

After computing the matching rate for each fingerprint-model pair, we aggregate these results across the models to compute statistics that describe each fingerprint’s matching behavior on the pirated model set $\mathcal{V}_f$ and the independent model set $\mathcal{I}_f$. For a given composite-sample fingerprint $\mathcal{T}_i$, we compute the mean matching rate $\mu_i^{\mathcal{V}}$ and standard deviation $\sigma_i^{\mathcal{V}}$ across all models in $\mathcal{V}_f$, and likewise obtain $\mu_i^{\mathcal{I}}$ and $\sigma_i^{\mathcal{I}}$ for $\mathcal{I}_f$. These results are subsequently used to select fingerprints and compute the fingerprint-specific thresholds.

\textbf{Step 2: Selecting composite-sample fingerprints.}
To refine the fingerprint set, we assess the discriminative capability of each fingerprint by quantifying the behavioral separation it induces between $\mathcal{V}_f$ and $\mathcal{I}_f$. Thus, we compute Cohen’s $d$ effect size for each $\mathcal{T}_i$ by $d_i = {(\mu_i^\mathcal{V} - \mu_i^\mathcal{I})}/{\sqrt{\frac{1}{2} \left( \left( \sigma_i^\mathcal{V} \right)^2 + \left( \sigma_i^\mathcal{I} \right)^2 \right)}}$.
Cohen’s $d$ effect size is employed to quantify the standardized separation between the two matching rate distributions of pirated and independently-trained models, capturing both the magnitude of behavioral difference and the statistical stability of that difference.
A larger $d_i$ indicates that a fingerprint $\mathcal{T}_i$ elicits more divergent behaviors between pirated and independent model sets, while maintaining higher consistency within each group, making it a stronger candidate for reliable ownership verification.
Based on $d_i$, we select the top-$K$ fingerprints with the highest $d_i$ values to construct a final fingerprint set $\mathcal{T}^* = \{\mathcal{T}^*_i\}_{i=1}^K$.

\renewcommand{\arraystretch}{0.85}  % reduce row height
\begin{table*}[htbp]
    \centering
    \caption{AUCs achieved by different fingerprinting approaches across datasets and protected model architectures.}
    % \caption{AUCs achieved by different approaches across DNN models and datasets.}
    \label{tab:auc_all}
    \vspace{-8pt}
    \resizebox{\textwidth}{!}{%
    \begin{tabular}{ccccccc}
        \toprule
        \textbf{Protected Model} & \textbf{Method} & \textbf{CIFAR-10}          & \textbf{CIFAR-100}         & \textbf{Fashion-MNIST}  & \textbf{MNIST} & \textbf{Tiny-ImageNet}\\
        \midrule
        \multirow{5}{*}{ResNet-18} 
        &IPGuard         & 0.675 $\pm$ 0.095          & 0.654 $\pm$ 0.097          & 0.721 $\pm$ 0.061       & 0.471 $\pm$ 0.089   &0.726 $\pm$ 0.099 \\
        &UAP             & 0.732 $\pm$ 0.014          & 0.761 $\pm$ 0.049          & 0.721 $\pm$ 0.061       & 0.789 $\pm$ 0.036   &0.812 $\pm$ 0.045\\
        &ADV-TRA         & 0.799 $\pm$ 0.003          & 0.806 $\pm$ 0.005          & 0.845 $\pm$ 0.010        & 0.753 $\pm$ 0.019  &0.767 $\pm$ 0.073\\
        &AKH             & 0.710 $\pm$ 0.052          & 0.785 $\pm$ 0.086          & 0.765 $\pm$ 0.042       & 0.820 $\pm$ 0.026   & 0.823 $\pm$ 0.043 \\
        &\textbf{IrisFP}  & \textbf{0.893 $\pm$ 0.015} & \textbf{0.916 $\pm$ 0.009} & \textbf{0.940 $\pm$ 0.031} & \textbf{0.854 $\pm$ 0.024} &\textbf{0.874 $\pm$ 0.052}\\

        \midrule
        \multirow{5}{*}{MobileNet-V2} 
        &IPGuard         & 0.821 $\pm$ 0.047          & 0.823 $\pm$ 0.021          & 0.607 $\pm$ 0.035       & 0.634 $\pm$ 0.010   &0.692 $\pm$ 0.019 \\
        &UAP             & 0.749 $\pm$ 0.028          & 0.836 $\pm$ 0.042          & 0.816 $\pm$ 0.019       & 0.743 $\pm$ 0.021   &0.806 $\pm$ 0.039\\
        &ADV-TRA         & 0.824 $\pm$ 0.051          & 0.795 $\pm$ 0.011          & 0.782 $\pm$ 0.028        & 0.720 $\pm$ 0.023  &0.877 $\pm$ 0.045\\
        &AKH             & 0.860 $\pm$ 0.067          & 0.867 $\pm$ 0.044          & 0.797 $\pm$ 0.054       & 0.805 $\pm$ 0.019   & 0.863 $\pm$ 0.034 \\
        &\textbf{IrisFP}  & \textbf{0.936 $\pm$ 0.011} & \textbf{0.937 $\pm$ 0.017} & \textbf{0.963 $\pm$ 0.017} & \textbf{0.876 $\pm$ 0.015} &\textbf{0.934 $\pm$ 0.023}\\

        \midrule
        \multirow{5}{*}{ViT-B/16} 
        &IPGuard         & --         & --              & --        & --     &0.778 $\pm$ 0.029 \\
        &UAP             & --           & --            & --         & --     &0.803 $\pm$ 0.042\\
        &ADV-TRA         & --           & --            & --         & --     &0.832 $\pm$ 0.019\\
        &AKH             & --           & --            & --        & --     &0.806 $\pm$ 0.010 \\
        &\textbf{IrisFP}  & --          & --            & --        & --   &\textbf{0.887 $\pm$ 0.036}\\
        
        \bottomrule
    \end{tabular}
    }
\vspace{-10pt}
\end{table*}

%%%%%%%%%%%%%%%%%%%%%%%%%%%%%%%%%%%%%%%%
\textbf{Step 3: Computing the fingerprint-specific threshold.}
To account for the varying discriminative power of individual fingerprints, we develop a fingerprint-specific thresholding strategy that assigns a specific threshold to each fingerprint. 
This design ensures that each fingerprint enforces a verification criterion consistent with its statistical separability between pirated and independently-trained models. The threshold defines the minimum matching rate required to assert a target model $f_t$ to be pirated.
Each threshold $\theta_i$ is computed based on the matching rate statistics of a fingerprint $\mathcal{T}^*_i$ across $\mathcal{V}_f$ and $\mathcal{I}_f$. More specifically, $\theta_i$ is calculated as a weighted average of their means, with weights inversely proportional to their standard deviations:
\begin{equation}
\theta_i = 
\begin{cases}
\frac{\mu_i^\mathcal{V}/\sigma_i^\mathcal{V} + \mu_i^\mathcal{I}/\sigma_i^\mathcal{I}}{1/\sigma_i^\mathcal{V} + 1/\sigma_i^\mathcal{I}}, & \text{if } \sigma_i^\mathcal{V}, \sigma_i^\mathcal{I} > 0, \\
\frac{\mu_i^\mathcal{V} + \mu_i^\mathcal{I}}{2}, & \text{otherwise}.
\end{cases}
\label{eq:weighted_threshold}
\end{equation}
This thresholding strategy assigns each fingerprint an individually optimized threshold, jointly considering the mean and standard deviation of matching-rate distributions to achieve maximal discriminative capability.

%______________________________________
\subsection{The Ownership Verification Process}
%______________________________________
To verify the ownership of a target model $f_{t}$, we assess how it responds to all samples in each composite-sample fingerprint within the fingerprint set $\mathcal{T}^*$. To this end, we design a two-step ownership verification scheme, comprising ownership matching and decision aggregation. The detailed design is presented below.

\paragraph{Step 1: Ownership matching.}
For each composite-sample fingerprint $\mathcal{T}^*_i \in \mathcal{T}^*$, its matching rate is computed by $MR(f_t, \mathcal{T}^*_i) = \frac{1}{T+1} \sum_{t=0}^{T} \mathbb{I}{[ f_t(\hat{x}_i^t) = \hat{y}_i^t ]}$.
Then, the resulting matching rate is compared with the corresponding fingerprint-specific threshold $\theta_i$, i.e., $SMR(f_t,\mathcal{T}^*_i) = \mathbb{I}\left[ MR(f_t,\mathcal{T}^*_i) \geq \theta_i \right]$,
where a result of ``1'' indicates that the target model's behavior on $\mathcal{T}^*_i$ is consistent with the protected model $f$, and ``0'' otherwise.

\paragraph{Step 2: Decision aggregation.} 
After obtaining the matching decision for each composite-sample fingerprint, a final ownership decision will be made. Since pirated models inherit decision behavior from the protected model $f$, they are expected to match the majority of fingerprints, whereas independently-trained models tend to diverge. We aggregate these decisions to produce a final verification score by $VS(f_t, \mathcal{T}^*) = \frac{1}{K} \sum_{i=1}^{K} SMR(f_t, \mathcal{T}^*_i)$,
which quantifies the overall behavioral alignment of the target model $f_t$ with $f$.
The target model is classified as pirated if $VS(f_t, \mathcal{T}^*) \geq \alpha$, where $\alpha$ is a predefined decision threshold; otherwise, it is deemed independently-trained. 

%% file: sec/4_experiments.tex
%%%%%%%%%%%%%%%%%%%%%%%%%%%%%%%%%
\section{Experiments}
\label{sec:experiments}
%%%%%%%%%%%%%%%%%%%%%%%%%%%%%%%%%

\subsection{Experimental Setting}
To evaluate the effectiveness of IrisFP, we conduct comprehensive experiments on different DNN models trained on five widely-used datasets: CIFAR-10 \cite{CIFAR10}, CIFAR-100 \cite{CIFAR100}, Fashion-MNIST \cite{FMNIST}, MNIST \cite{MNIST}, and Tiny-ImageNet \cite{TinyImageNet}.
We compare IrisFP with four representative fingerprinting methods: IPGuard \cite{IPIP21}, UAP \cite{FDNN22}, ADV-TRA \cite{UWSDW24}, and AKH\cite{QURD25}. We also evaluate the robustness of IrisFP against model modification attacks by considering six types of attacks, including fine-tuning (FT), pruning (PR), knowledge distillation (KD), adversarial training (AT), prune-then-tune (PFT), and noise injection-then-fine-tuning (NFT). 
To ensur fair comparisons, we query 200 times in each experiment, i.e., 40 composite fingerprints with 5 samples each for our proposed IrisFP and 200 fingerprints for all baseline methods.
In the following, we will summarize protected models, reference model sets, testing model set, and evaluation metrics, and their detailed settings can be found in Sec.~\ref{sec:appendix_expsetting} in the Supplementary Material. 

\textbf{Protected Model.} We consider three different architectures, including ResNet-18, MobileNet-V2, and ViT-B/16. 
For ResNet-18 and MobileNet-V2, each protected model is trained from scratch across its corresponding different dataset, while for ViT-B/16, its protected model is fine-tuned from a pretrained ViT-B/16 model on Tiny-ImageNet.

\textbf{Reference Model Sets.} We construct two reference model sets: one for pirated models and the other for independently-trained models. Specifically, for each protected model, its corresponding pirated model set is constructed by modifying the protected model through three model modification techniques, including fine-tuning (FT), knowledge distillation (KD), and adversarial training (AT). Each attack type produces three pirated variants, leading to 9 pirated variants in the pirated model set. On the other hand, our independent model set is constructed by using three architectures: ResNet-18, MobileNet-V2, and DenseNet-121. The models in this set are independently trained from scratch with different seeds and hyperparameters from training the protected models and testing models. Each architecture will be instantiated with 3 different random seeds to create 3 independently-trained models, resulting in a total of 9 models in the independent model set.

\textbf{Testing Model Set.}
We evaluate IrisFP using a testing model set comprising both pirated and independently-trained models. The pirated models are derived by applying six attacks (i.e., FT, PR, KD, AT, PFT, and NFT) to each protected model. For each attack type, we generate 20 variants using different random seeds and hyperparameters, resulting in 120 pirated models. The independently-trained models are built from scratch using six architectures: ResNet-18, ResNet-50, MobileNet-V2, MobileNet-V3 Large, EfficientNet-B2, and DenseNet-121. For each architecture, 20 models are trained with distinct seeds and hyperparameters, without any access to the protected models or the models in the reference model set, yielding 120 independently-trained models.

\textbf{Metrics.}
We use the Area Under the ROC (Receiver Operating Characteristic) Curve (AUC) as the main metric to evaluate the performance of our proposed fingerprinting method. AUC quantifies the probability that a randomly selected pirated model receives a higher matching score than a randomly selected independently-trained model. Higher AUC values reflect stronger discriminative capability. All AUC results are reported as the mean and standard deviation over five independent runs.

%%%%%%%%%%%%%%%%%%%%%%%%%%%%%%%%
\subsection{Main Performance}
%%%%%%%%%%%%%%%%%%%%%%%%%%%%%%%%

\setlength{\tabcolsep}{1mm}
\renewcommand{\arraystretch}{0.7}  % reduce row height
\begin{table*}[h]
    \centering
    \caption{AUCs under six model modification attacks across datasets and methods on the protected model with ResNet-18 architecture.}
\vspace{-8pt}
\label{tab:auc_diff_attacks}
    \resizebox{\textwidth}{!}{
    \begin{tabular}{llcccccc}
        \toprule
        \textbf{Dataset} & \textbf{Method} & \textbf{FT} & \textbf{PR} & \textbf{KD} & \textbf{AT} & \textbf{PFT} & \textbf{NFT} \\
        \midrule
        \multirow{4}{*}{CIFAR-10} 
        
        & IPGuard     & 0.656 $\pm$ 0.101 & 0.997 $\pm$ 0.005 & 0.515 $\pm$ 0.082 & 0.511 $\pm$ 0.265 & 0.687 $\pm$ 0.144 & 0.724 $\pm$ 0.142 \\
        & UAP         & 0.809 $\pm$ 0.011 & 0.998 $\pm$ 0.005 & 0.722 $\pm$ 0.013 & 0.346 $\pm$ 0.017 & 0.856 $\pm$ 0.016 & 0.868 $\pm$ 0.018 \\
        & ADV-TRA     & \textbf{1.000 $\pm$ 0.000} & \textbf{1.000 $\pm$ 0.000} & \textbf{0.805 $\pm$ 0.009} & 0.025 $\pm$ 0.006 & 0.959 $\pm$ 0.012 & 0.962 $\pm$ 0.009 \\
        & AKH & 0.921 $\pm$ 0.005 & 0.876 $\pm$ 0.020 & 0.621 $\pm$ 0.042 & 0.531 $\pm$ 0.108 & 0.701 $\pm$ 0.029 & 0.733 $\pm$ 0.066 \\
        & \textbf{IrisFP}       & 0.954 $\pm$ 0.046 & \textbf{1.000 $\pm$ 0.000} & 0.616 $\pm$ 0.068 & \textbf{0.929 $\pm$ 0.042} & \textbf{0.965 $\pm$ 0.009} & \textbf{0.968 $\pm$ 0.014} \\
        \midrule
        \multirow{4}{*}{CIFAR-100} 
        
        & IPGuard     & 0.617 $\pm$ 0.137 & 0.998 $\pm$ 0.003 & 0.593 $\pm$ 0.141 & 0.621 $\pm$ 0.106 & 0.576 $\pm$ 0.104 & 0.584 $\pm$ 0.129 \\
        & UAP         & 0.764 $\pm$ 0.069 & 0.900 $\pm$ 0.002 & 0.709 $\pm$ 0.071 & 0.677 $\pm$ 0.053 & 0.752 $\pm$ 0.052 & 0.770 $\pm$ 0.065 \\
        & ADV-TRA     & 0.863 $\pm$ 0.010 & \textbf{1.000 $\pm$ 0.000} & 0.899 $\pm$ 0.007 & 0.338 $\pm$ 0.008 & 0.864 $\pm$ 0.014 & 0.863 $\pm$ 0.012 \\
        & AKH & 0.813 $\pm$ 0.076 & 0.670 $\pm$ 0.185 & 0.858 $\pm$ 0.061 & 0.221 $\pm$ 0.076 & 0.671 $\pm$ 0.050 & 0.775 $\pm$ 0.090 \\
        & \textbf{IrisFP}       & \textbf{0.953 $\pm$ 0.011} & \textbf{1.000 $\pm$ 0.000} & \textbf{0.927 $\pm$ 0.006} & \textbf{0.758 $\pm$ 0.031} & \textbf{0.933 $\pm$ 0.009} & \textbf{0.928 $\pm$ 0.019} \\
        \midrule
        \multirow{4}{*}{Fashion-MNIST} 
       
        & IPGuard     & 0.688 $\pm$ 0.139 & 0.990 $\pm$ 0.001 & 0.592 $\pm$ 0.096 & 0.612 $\pm$ 0.115 & 0.632 $\pm$ 0.187 & 0.636 $\pm$ 0.216 \\
        & UAP         & 0.809 $\pm$ 0.055 & 0.869 $\pm$ 0.045 & 0.749 $\pm$ 0.012 & 0.716 $\pm$ 0.018 & 0.797 $\pm$ 0.061 & 0.781 $\pm$ 0.045 \\
        & ADV-TRA     & 0.980 $\pm$ 0.000 & 0.990 $\pm$ 0.000 & 0.830 $\pm$ 0.004 & 0.158 $\pm$ 0.008 & 0.981 $\pm$ 0.001 & 0.980 $\pm$ 0.000 \\
        & AKH & 0.855 $\pm$ 0.118 & 0.990 $\pm$ 0.000 & 0.844 $\pm$ 0.006 & 0.203 $\pm$ 0.038 & 0.759 $\pm$ 0.078 & 0.737 $\pm$ 0.091 \\
        & \textbf{IrisFP}       & \textbf{0.982 $\pm$ 0.002} & \textbf{0.992 $\pm$ 0.000} & \textbf{0.853 $\pm$ 0.037} & \textbf{0.816 $\pm$ 0.190} & \textbf{0.982 $\pm$ 0.003} & \textbf{0.983 $\pm$ 0.002} \\
        \midrule
        \multirow{4}{*}{MNIST} 
        
        & IPGuard     & 0.378 $\pm$ 0.022 & 0.733 $\pm$ 0.030 & {0.624} $\pm$ 0.108 & 0.067 $\pm$ 0.065 & 0.432 $\pm$ 0.045 & 0.456 $\pm$ 0.122 \\
        & UAP         & 0.966 $\pm$ 0.009 & 0.630 $\pm$ 0.011 & 0.455 $\pm$ 0.054 & \textbf{0.946 $\pm$ 0.088} & 0.892 $\pm$ 0.007 & 0.959 $\pm$ 0.005 \\
        & ADV-TRA     & 0.965 $\pm$ 0.002 & 0.753 $\pm$ 0.028 & 0.485 $\pm$ 0.023 & 0.475 $\pm$ 0.019 & 0.975 $\pm$ 0.003 & 0.976 $\pm$ 0.001\\
        & AKH & 0.980 $\pm$ 0.000 & 0.797 $\pm$ 0.013 & \textbf{0.745 $\pm$ 0.011} & 0.837 $\pm$ 0.069 & 0.915 $\pm$ 0.010 & 0.920 $\pm$ 0.030 \\
        & \textbf{IrisFP}       & \textbf{0.985 $\pm$ 0.001} & \textbf{0.966 $\pm$ 0.007} & 0.467 $\pm$ 0.091 & 0.704 $\pm$ 0.025 & \textbf{0.983 $\pm$ 0.002} & \textbf{0.985 $\pm$ 0.001} \\
         \midrule
        \multirow{4}{*}{Tiny-ImageNet} 
        
        & IPGuard     & 0.958 $\pm$ 0.044 & 0.754 $\pm$ 0.129 & 0.525 $\pm$ 0.136 & 0.316 $\pm$ 0.118 & 0.867 $\pm$ 0.133 & 0.954 $\pm$ 0.034 \\
        & UAP         & 0.953 $\pm$ 0.038 & 0.981 $\pm$ 0.001 & 0.455 $\pm$ 0.089 & 0.411 $\pm$ 0.114 & 0.967 $\pm$ 0.012 & 0.952 $\pm$ 0.016 \\
        & ADV-TRA     & 0.970 $\pm$ 0.040 & 0.917 $\pm$ 0.056 & 0.466 $\pm$ 0.069 & 0.173 $\pm$ 0.143 & 0.980 $\pm$ 0.064 & 0.948 $\pm$ 0.036 \\
        & AKH & 0.937 $\pm$ 0.040  & 0.875 $\pm$ 0.019 & {0.326} $\pm$ 0.055 & 0.488 $\pm$ 0.099 & 0.988 $\pm$ 0.024 & 0.944 $\pm$ 0.051 \\
        & \textbf{IrisFP}       & \textbf{0.986 $\pm$ 0.032} & \textbf{1.000 $\pm$ 0.000} & \textbf{0.661 $\pm$ 0.108} & \textbf{0.545 $\pm$ 0.106} & \textbf{0.992 $\pm$ 0.015} & \textbf{0.977 $\pm$ 0.027} \\
        \bottomrule
        
    \end{tabular}
    }
\vspace{-10pt}
\end{table*}

%%%%%%%%%%%%%%%%%%%%%%%%%%%%%%%%%%%
\subsubsection{The Effectiveness of IrisFP}
%%%%%%%%%%%%%%%%%%%%%%%%%%%%%%%%%%%

Table~\ref{tab:auc_all} shows the superior effectiveness of IrisFP by reporting the AUCs achieved by IrisFP and four representative baselines (IPGuard, UAP, ADV-TRA, and AKH) across five benchmark datasets—CIFAR-10, CIFAR-100, Fashion-MNIST, MNIST, and Tiny-ImageNet. To ensure a comprehensive evaluation, we train two protected models using ResNet-18 and MobileNet-V2 architectures on each of the five datasets. In addition, we incorporate a larger and more complex architecture by training a protected model using ViT-B/16 on Tiny-ImageNet. From the table, we can observe that IrisFP consistently achieves the highest AUC across the models and datasets. For example, on CIFAR-100, IrisFP achieves an AUC of 0.916 when fingerprinting ResNet-18 and an AUC of 0.937 when fingerprinting MobileNet-V2, achieving the improvements by 13.7\% and 8.1\%, respectively, over the best-performing baselines. It is noteworthy that IrisFP maintains the highest AUC of 0.887 when fingerprinting ViT-B/16 trained on Tiny-ImageNet, demonstrating that IrisFP can achieve reliable ownership verification on larger and more complex models.

\vspace{-5pt}
\begin{figure}[htbp]
    \centering
    \includegraphics[width=0.9\linewidth]{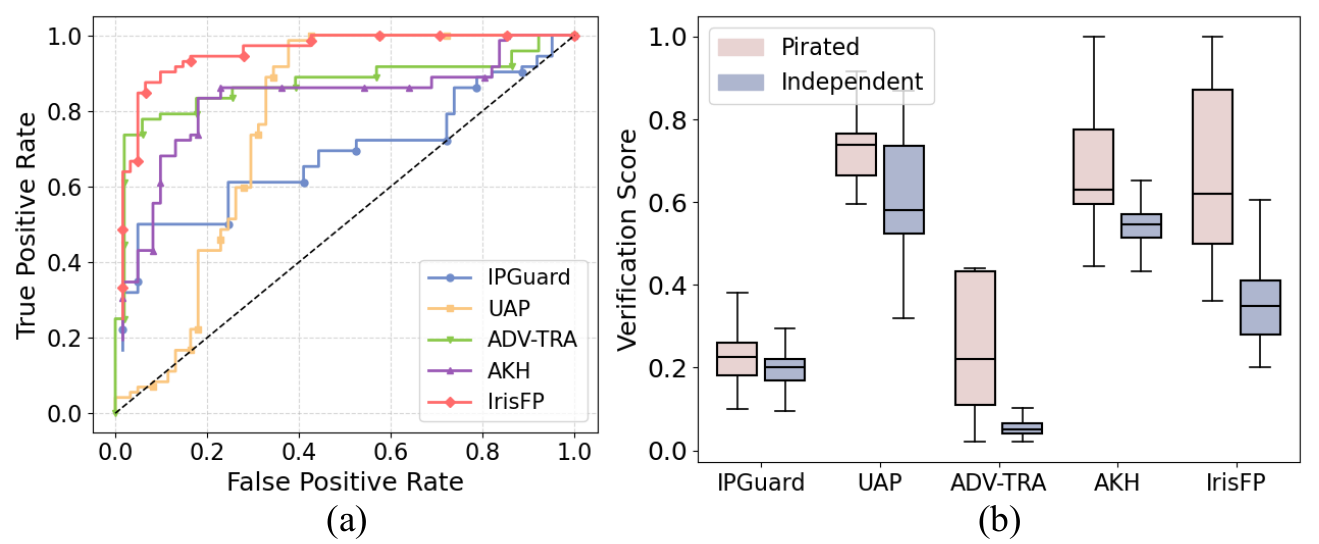}
   % \vspace{-0.2in}
    \vspace{-8pt}
    \caption{(a) The ROC curve and (b) the distribution of verification score on the protected model with ResNet-18 architecture trained on Fashion-MNIST.}
    \vspace{-5pt}
\label{fig:roc_box}
\end{figure} 

\begin{figure*}[htbp]
    \centering
    \includegraphics[width=0.9\linewidth]{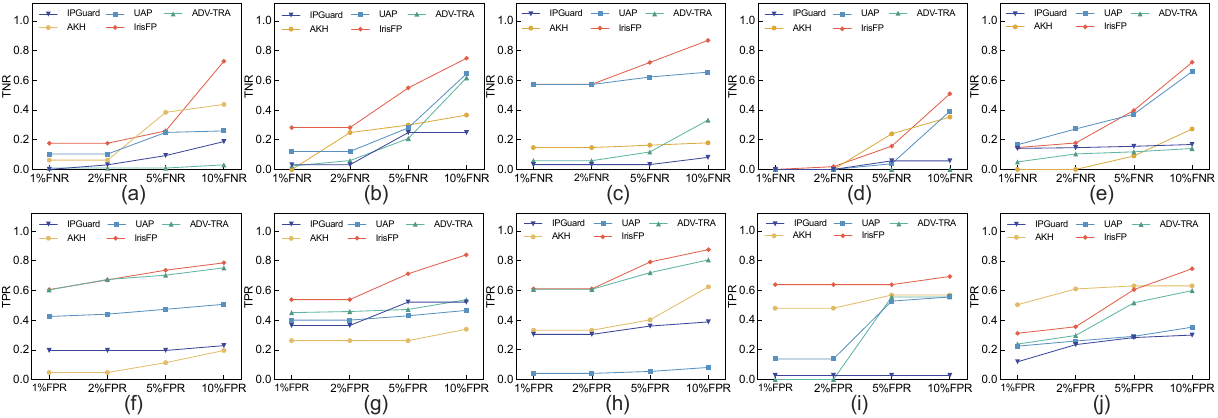}
   % \vspace{-0.2in}
    \vspace{-5pt}
    \caption{TNR-FNR and TPR-FPR curves on the protected models with ResNet-18 architecture trained on CIFAR-10 ((a) and (f)), CIFAR-100 ((b) and (g)), Fashion-MNIST ((c) and (h)), MNIST ((d) and (i)), and Tiny-ImageNet ((e) and (j)).}
    \vspace{-10pt}
\label{fig:robust_unique}
\end{figure*} 
\vspace{-5pt}

To further validate IrisFP's discriminative power, Figure~\ref{fig:roc_box} visualizes the ROC curves and the distribution of verification scores. These plots provide intuitive insights into how well each method distinguishes pirated models from independently-trained ones. Figure~\ref{fig:roc_box}(a) shows the ROC curves for all five methods, illustrating the trade-off between TPR and FPR. 
IrisFP clearly demonstrates the strongest discriminative capability, with its curve close to the top-left corner, outperforming all baselines. 
In contrast, IPGuard's and UAP's curves lie near the diagonal, indicating weak uniqueness achieved by them, while ADV-TRA and AKH perform better but still fall short of IrisFP. 
Figure~\ref{fig:roc_box}(b) presents two distributions of fingerprint verification scores achieved by each fingerprinting method under both pirated and independently-trained models.
A wider separation between the two distributions reflects a stronger discriminative capability. 
As shown in this figure, IrisFP achieves a clear separation, while baselines exhibit greater distributional overlap, reflecting their limited ability to differentiate pirated models from independently-trained ones.

%%%%%%%%%%%%%%%%%%%%%%%%%%%%%%%%%%%%%%%%%%%%%%%%%%%
\subsubsection{Evaluation of IrisFP’s Uniqueness and Robustness}
Figure~\ref{fig:robust_unique} demonstrates the uniqueness and robustness achieved by IrisFP. The evaluation is performed under varying FNR and FPR levels, specifically 1\%, 2\%, 5\%, and 10\%. We can observe from the figure that IrisFP consistently outperforms all baselines, achieving the highest TNR (as shown by (a)-(e)) and the highest TPR (as shown by (f)-(j)), thereby demonstrating superior uniqueness and robustness. Among the baselines, ADV-TRA exhibits strong robustness—achieving high TPR by effectively identifying pirated models—but suffers from low TNR, indicating poor uniqueness due to frequent misidentification of independently-trained models. In contrast, UAP maintains relatively stable TNR, indicating better uniqueness, but struggles with lower TPR, especially on Fashion-MNIST, revealing limited robustness. Moreover,
AKH shows inconsistent behavior, performing well in either robustness or uniqueness but not both. Additionally,
IPGuard performs the worst, failing to reliably distinguish pirated models from independently-trained ones under any setting. These results demonstrate that IrisFP is the only method achieving high robustness and uniqueness simultaneously, making it a reliable model fingerprinting solution.

%%%%%%%%%%%%%%%%%%%%%%%%%%%%%%%%%%%%%%%%%%%%%%%%%%%
\subsubsection{Robustness to Model Modification Attacks}
Table~\ref{tab:auc_diff_attacks} shows the effectiveness of IrisFP against model modifications that aim to remove or distort the original fingerprints.
Specifically, we compare IrisFP with four representative baselines using the protected model with ResNet-18 architecture trained across five datasets—CIFAR-10, CIFAR-100, Fashion-MNIST, MNIST, and Tiny-ImageNet—under six representative attacks: FT, PR, KD, AT, NFT, and PFT. The comparison is based on the AUCs achieved under each setting. We can observe that IrisFP consistently achieves the best performance under CIFAR-100, Fashion-MNIST, and Tiny-ImageNet across all six attacks. For instance, IrisFP attains AUCs of 0.953 (FT), 0.927 (KD), and 0.758 (AT) for CIFAR-100, and 0.982 (FT), 0.853 (KD), and 0.816 (AT) for Fashion-MNIST, substantially outperforming the baselines. For CIFAR-10, IrisFP achieves the highest AUCs under four attacks (PR, AT, PFT, NFT) and slightly lower AUC under FT and KD.
These observations demonstrate IrisFP's strong robustness to a wide range of model modification attacks.

\vspace{-5pt}
\begin{figure}[htbp]
    \centering    \includegraphics[width=0.99\linewidth]{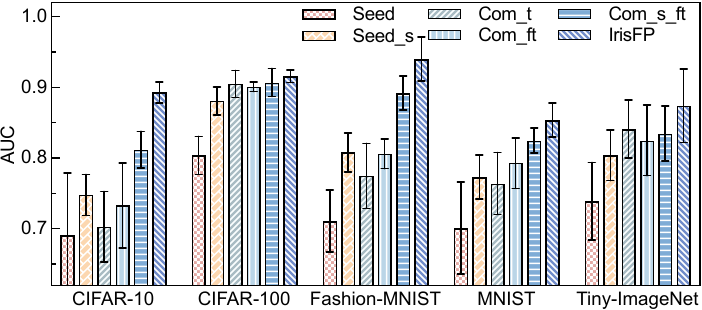}
    % \vspace{-0.2in}
    \vspace{-5pt}
    \caption{AUCs under different fingerprint configurations on the protected model with ResNet-18 architecture.}
\label{fig:abl_construction}
\vspace{-8pt}
\end{figure}
\vspace{-5pt}

\subsection{Ablation Studies}

We conduct ablation studies to evaluate the effectiveness of each component in IrisFP, including fingerprint seed initialization, composite-sample fingerprint generation, fingerprint selection, and fingerprint-specific threshold: 
% , as shown in Table~\ref{tab:abl_config}:
\begin{itemize}[left=0pt]
    \item Seed: Directly using initial fingerprint seeds without further processing.
    
    \item Seed\_s: Applying fingerprint selection to fingerprint seeds. 
    %without constructing unified fingerprints and applying fingerprint-specific thresholding.
    
    \item Com\_ft: Extending fingerprint seeds into composite-sample fingerprints and using a fixed global threshold.  
    \item Com\_s\_ft: Extending fingerprint seeds into composite-sample fingerprints, using a fixed global threshold, and applying fingerprint selection.
    
    \item Com\_t: Extending fingerprint seeds into composite-sample fingerprints and applying fingerprint-specific thresholds, without fingerprint selection.

\end{itemize}

Figure~\ref{fig:abl_construction} illustrates the AUCs across five datasets under various configurations. 
Specifically, Seed alone yields the worst verification performance across all datasets, as it only reflects the initial fingerprinting step without other components. Seed\_s leads to a clear improvement (e.g., from 0.691 to 0.748 on CIFAR-10), demonstrating the necessity of retaining only the most discriminative fingerprints.
Furthermore, extending fingerprint seeds into composite-sample fingerprints (Com\_ft) enhances uniqueness by capturing richer decision-region behavior through class-diverse outputs. Building upon this, applying fingerprint selection (Com\_s\_ft) further strengthens discriminative capability by filtering out less effective fingerprints. More importantly, using fingerprint-specific thresholds also results in performance improvement, for example, improving the AUC from 0.812 (Com\_s\_ft) to 0.893 (IrisFP) on CIFAR-10, demonstrating the advantage of adapting to a specific threshold for each composite-sample fingerprint.
In addition, IrisFP consistently outperforms all ablated variants across datasets. These results highlight the complementary strengths of each component and the necessity of integrating all of them in our design for reliable ownership verification.

Further hyperparameter analysis and additional experimental results are presented in the supplementary material.

%% file: sec/5_conclusion.tex
\vspace{-5pt}
%%%%%%%%%%%%%%%%%%%%%%%%%%%%%%%
\section{Conclusions}
\label{sec:conclusion}
%%%%%%%%%%%%%%%%%%%%%%%%%%%%%%%
\vspace{-5pt}
This paper studies the problem of enhancing both uniqueness and robustness for adversarial-example-based model fingerprinting. To address this issue, we propose IrisFP, a new fingerprinting method that leverages multi-boundary characteristics, multi-sample behavior, and fingerprint quality assessment. Specifically, IrisFP crafts fingerprint seeds positioned near the intersection of multi-decision boundaries to increase prediction margins. It then applies subtle and diverse perturbations to each seed to generate multiple variants, forming a composite-sample fingerprint. The resulting fingerprints are refined using a discriminative selection strategy to retain only those with the strongest separability. For ownership verification, IrisFP employs a two-step process comprising ownership matching and decision aggregation, tailored to the structure of composite-sample fingerprints. Extensive experiments validate the effectiveness of IrisFP, with results consistently showing superior performance over state-of-the-art methods.

%% file: sec/6_appendix.tex
\clearpage
\setcounter{page}{1}
\maketitlesupplementary

%%%%%%%%%%%%%%%%%%%%%%%%%%%%%%%%%%%%%%%%%%%%
% \section{Overview}
\label{sec:appendix}
%%%%%%%%%%%%%%%%%%%%%%%%%%%%%%%%%%%%%%%%%%%%

%%%%%%%%%%%%%%%%%%%%%%%%%%%%%%%%%
\section{Threat Model}
%%%%%%%%%%%%%%%%%%%%%%%%%%%%%%%%%

We consider a typical model fingerprinting scenario involving two entities: a model owner and an adversary. 
Specifically, the model owner first produces a protected model $f$ and then exploits adversarial example techniques to generate a set of fingerprints, which are kept confidential, to protect the IP of $f$. On the other hand, the adversary first obtains an unauthorized copy of $f$, for example, via white-box access or black-box extraction, and subsequently modifies it using model modification techniques, such as fine-tuning (FT), pruning (PR), adversarial training (AT), knowledge distillation (KD), etc., to produce a variant $f_p$. The adversary then deploys $f_p$ in a black-box setting, making it accessible to the public through APIs. 

Suppose that the model owner identifies a black-box model $f_t$ as a suspicious target. The owner's goal is to determine whether $f_t$ is derived from the protected model $f$. To perform ownership verification, the owner queries $f_t$ using the generated fingerprints and compares the returned outputs with the expected fingerprint outputs. If the number of matches exceeds a predefined threshold, the target model is deemed to be an infringing model.

%%%%%%%%%%%%%%%%%%%%%%%%%
\section{Additional Results}
%%%%%%%%%%%%%%%%%%%%%%

Tables~\ref{tab:auc_diff_attacks_mobilenet} and~\ref{tab:auc_vit_tinyimagenet} present additional results on the robustness of the protected models using the MobileNet-V2 and ViT-B/16 architectures, respectively. For MobileNet-V2, we evaluate IrisFP against four representative baselines across five datasets—CIFAR-10, CIFAR-100, Fashion-MNIST, MNIST, and Tiny-ImageNet—under six model modification attacks: FT, PR, KD, AT, NFT, and PFT. IrisFP remains consistently robust across datasets, achieving the highest AUCs on CIFAR-100, Fashion-MNIST, and MNIST under all attacks, and leading on Tiny-ImageNet for five out of six attacks (FT, KD, AT, PFT, NFT), with only a slight drop on PR. On CIFAR-10, IrisFP attains the highest AUCs under PR, AT, and PFT, while remaining competitive on the remaining attacks. For the protected model with ViT-B/16 architecture trained on Tiny-ImageNet, IrisFP achieves the highest AUCs across all six attacks, demonstrating strong robustness on transformer-based architectures as well. Together with the results on ResNet-18 in the main text (i.e., Table~\ref{tab:auc_diff_attacks}), these findings confirm that IrisFP maintains high robustness against model modification attacks across diverse model architectures---lightweight and high-capacity convolutional models as well as modern vision transformers.

\setlength{\tabcolsep}{1mm}
\renewcommand{\arraystretch}{0.7}  % reduce row height
\begin{table*}[h]
    \centering
    \caption{AUCs under six attacks across datasets and methods on the protected model with MobileNet-V2 architecture.}
\vspace{-8pt}
\label{tab:auc_diff_attacks_mobilenet}
    \resizebox{\textwidth}{!}{
    \begin{tabular}{llcccccc}
        \toprule
        \textbf{Dataset} & \textbf{Method} & \textbf{FT} & \textbf{PR} & \textbf{KD} & \textbf{AT} & \textbf{PFT} & \textbf{NFT} \\
        \midrule
        \multirow{4}*{CIFAR-10} 
        & IPGuard     & 0.951 $\pm$ 0.022 & 0.994 $\pm$ 0.000 & 0.622 $\pm$ 0.048 & 0.481 $\pm$ 0.127 & 0.942 $\pm$ 0.029 & 0.934 $\pm$ 0.061 \\
        & UAP         & 0.891 $\pm$ 0.011 & 0.915 $\pm$ 0.003 & 0.526 $\pm$ 0.052 & 0.412 $\pm$ 0.098 & 0.892 $\pm$ 0.031 & 0.880 $\pm$ 0.009 \\
        & ADV-TRA     & \textbf{0.992 $\pm$ 0.025} & {0.995 $\pm$ 0.001} & 0.780 $\pm$ 0.068 & 0.214 $\pm$ 0.045 & {0.992 $\pm$ 0.040} & \textbf{0.992 $\pm$ 0.063} \\
        & AKH         & 0.896 $\pm$ 0.034 & 0.904 $\pm$ 0.010 & \textbf{0.885 $\pm$ 0.071} & 0.727 $\pm$ 0.052 & 0.935 $\pm$ 0.073 & 0.921 $\pm$ 0.084 \\
        & \textbf{IrisFP}   & 0.981 $\pm$ 0.006 & \textbf{0.997 $\pm$ 0.000} & 0.712 $\pm$ 0.055 & \textbf{0.978 $\pm$ 0.015} & \textbf{0.995 $\pm$ 0.003} & 0.982 $\pm$ 0.005 \\
        \midrule
        \multirow{4}*{CIFAR-100} 
        & IPGuard     & 0.981 $\pm$ 0.002 & 0.983 $\pm$ 0.005 & 0.741 $\pm$ 0.036 & 0.208 $\pm$ 0.087 & 0.982 $\pm$ 0.000 & 0.982 $\pm$ 0.001 \\
        & UAP         & 0.979 $\pm$ 0.010 & 0.987 $\pm$ 0.003 & 0.774 $\pm$ 0.079 & 0.723 $\pm$ 0.036 & 0.904 $\pm$ 0.056 & 0.929 $\pm$ 0.014 \\
        & ADV-TRA     & 0.963 $\pm$ 0.012 & 0.981 $\pm$ 0.001 & 0.701 $\pm$ 0.027 & 0.632 $\pm$ 0.018 & 0.885 $\pm$ 0.003 & 0.932 $\pm$ 0.009 \\
        & AKH         & 0.958 $\pm$ 0.007 & 0.979 $\pm$ 0.028 & 0.783 $\pm$ 0.068 & 0.656 $\pm$ 0.075 & 0.951 $\pm$ 0.057 & 0.923 $\pm$ 0.005 \\
        & \textbf{IrisFP}   & \textbf{0.983 $\pm$ 0.002} & \textbf{0.991 $\pm$ 0.001} & \textbf{0.811 $\pm$ 0.051} & \textbf{0.886 $\pm$ 0.055} & \textbf{0.983 $\pm$ 0.000} & \textbf{0.983 $\pm$ 0.000} \\
        \midrule
        \multirow{4}*{Fashion-MNIST} 
        & IPGuard     & 0.587 $\pm$ 0.027 & 0.945 $\pm$ 0.046 & 0.518 $\pm$ 0.118 & 0.550 $\pm$ 0.090 & 0.555 $\pm$ 0.016 & 0.543 $\pm$ 0.103 \\
        & UAP         & 0.897 $\pm$ 0.008 & 0.977 $\pm$ 0.007 & 0.658 $\pm$ 0.076 & 0.735 $\pm$ 0.031 & 0.971 $\pm$ 0.005 & 0.886 $\pm$ 0.013 \\
        & ADV-TRA     & 0.963 $\pm$ 0.032 & 0.964 $\pm$ 0.012 & 0.721 $\pm$ 0.058 & 0.871 $\pm$ 0.036 & 0.898 $\pm$ 0.007 & 0.723 $\pm$ 0.037 \\
        & AKH         & 0.879 $\pm$ 0.051 & 0.884 $\pm$ 0.038 & 0.821 $\pm$ 0.079 & 0.832 $\pm$ 0.040 & 0.855 $\pm$ 0.078 & 0.796 $\pm$ 0.016 \\
        & \textbf{IrisFP}   & \textbf{0.975 $\pm$ 0.012} & \textbf{0.991 $\pm$ 0.000} & \textbf{0.860 $\pm$ 0.109} & \textbf{0.960 $\pm$ 0.017} & \textbf{0.979 $\pm$ 0.006} & \textbf{0.978 $\pm$ 0.010} \\
        \midrule
        \multirow{4}*{MNIST} 
        & IPGuard     & 0.690 $\pm$ 0.007 & 0.710 $\pm$ 0.030 & 0.656 $\pm$ 0.020 & 0.560 $\pm$ 0.021 & 0.535 $\pm$ 0.003 & 0.757 $\pm$ 0.018 \\
        & UAP         & 0.767 $\pm$ 0.058 & 0.910 $\pm$ 0.000 & 0.668 $\pm$ 0.057 & 0.544 $\pm$ 0.011 & 0.672 $\pm$ 0.060 & 0.693 $\pm$ 0.022 \\
        & ADV-TRA     & 0.821 $\pm$ 0.039 & 0.896 $\pm$ 0.013 & 0.642 $\pm$ 0.033 & 0.557 $\pm$ 0.032 & 0.634 $\pm$ 0.014 & 0.703 $\pm$ 0.021 \\
        & AKH         & 0.871 $\pm$ 0.011 & 0.921 $\pm$ 0.015 & 0.569 $\pm$ 0.023 & 0.540 $\pm$ 0.052 & 0.885 $\pm$ 0.021 & 0.836 $\pm$ 0.009 \\
        & \textbf{IrisFP}   & \textbf{0.983 $\pm$ 0.000} & \textbf{0.960 $\pm$ 0.014} & \textbf{0.683 $\pm$ 0.041} & \textbf{0.577 $\pm$ 0.107} & \textbf{0.981 $\pm$ 0.001} & \textbf{0.981 $\pm$ 0.003} \\
        \midrule
        \multirow{4}*{Tiny-ImageNet} 
        & IPGuard     & 0.466 $\pm$ 0.086 & \textbf{0.996 $\pm$ 0.005} & 0.786 $\pm$ 0.006 & 0.538 $\pm$ 0.237 & 0.478 $\pm$ 0.074 & 0.475 $\pm$ 0.056 \\
        & UAP         & 0.896 $\pm$ 0.014 & 0.985 $\pm$ 0.003 & 0.799 $\pm$ 0.035 & 0.633 $\pm$ 0.097 & 0.872 $\pm$ 0.036 & 0.843 $\pm$ 0.038 \\
        & ADV-TRA     & 0.923 $\pm$ 0.013 & 0.932 $\pm$ 0.011 & 0.805 $\pm$ 0.077 & 0.836 $\pm$ 0.040 & 0.840 $\pm$ 0.087 & 0.848 $\pm$ 0.029 \\
        & AKH         & 0.850 $\pm$ 0.037 & 0.901 $\pm$ 0.023 & 0.820 $\pm$ 0.052 & 0.809 $\pm$ 0.014 & 0.843 $\pm$ 0.038 & 0.859 $\pm$ 0.024 \\
        & \textbf{IrisFP}   & \textbf{0.979 $\pm$ 0.011} & 0.956 $\pm$ 0.043 & \textbf{0.883 $\pm$ 0.096} & \textbf{0.863 $\pm$ 0.160} & \textbf{0.995 $\pm$ 0.001} & \textbf{0.963 $\pm$ 0.009} \\
        \bottomrule
    \end{tabular}
    }
% \vspace{-10pt}
\end{table*}

\setlength{\tabcolsep}{1mm}
\renewcommand{\arraystretch}{0.7}
\begin{table*}[h]
    \centering
    \caption{AUCs under six model modification attacks on the protected model with ViT-B/16 architecture.}
\vspace{-8pt}
\label{tab:auc_vit_tinyimagenet}
\resizebox{\textwidth}{!}{
\begin{tabular}{llcccccc}
\toprule
\textbf{Dataset} & \textbf{Method} & \textbf{FT} & \textbf{PR} & \textbf{KD} & \textbf{AT} & \textbf{PFT} & \textbf{NFT} \\
\midrule
\multirow{4}*{Tiny-ImageNet}
    & IPGuard  & 0.923 $\pm$ 0.006 & 0.692 $\pm$ 0.041 & 0.585 $\pm$ 0.153 & 0.606 $\pm$ 0.239 & 0.864 $\pm$ 0.036 & 0.901 $\pm$ 0.012 \\
    & UAP      & 0.979 $\pm$ 0.011 & 0.705 $\pm$ 0.030 & 0.593 $\pm$ 0.086 & 0.762 $\pm$ 0.071 & 0.731 $\pm$ 0.045 & 0.852 $\pm$ 0.022 \\
    & ADV-TRA  & 0.973 $\pm$ 0.022 & 0.765 $\pm$ 0.010 & 0.409 $\pm$ 0.023 & 0.901 $\pm$ 0.038 & 0.801 $\pm$ 0.008 & 0.915 $\pm$ 0.036 \\
    & AKH      & 0.938 $\pm$ 0.006 & 0.679 $\pm$ 0.007 & 0.539 $\pm$ 0.017 & 0.853 $\pm$ 0.014 & 0.738 $\pm$ 0.012 & 0.878 $\pm$ 0.010 \\
    & \textbf{IrisFP} 
               & \textbf{0.960 $\pm$ 0.007} 
               & \textbf{0.829 $\pm$ 0.024} 
               & \textbf{0.656 $\pm$ 0.045} 
               & \textbf{0.939 $\pm$ 0.060} 
               & \textbf{0.766 $\pm$ 0.114} 
               & \textbf{0.948 $\pm$ 0.003} \\
\bottomrule
\end{tabular}
}
% \vspace{-10pt}
\end{table*}

%%%%%%%%%%%%%%%%%%%%%%%%%%%%%%%%%%%%%%%%%%%%
\section{Impact of Hyperparameters}
\label{sec:appendix_impactparameters}
%%%%%%%%%%%%%%%%%%%%%%%%%%%%%%%%%%%%%%%%%%%%
To ensure a comprehensive evaluation, we consider three protected models in the following experiments: a ResNet-18 trained on CIFAR-100, a MobileNet-V2 trained on Fashion-MNIST, and a ViT-B/16 trained on Tiny-ImageNet.

\subsection{Impact of combination of $K$ and $T$}
\begin{figure}[htbp]
    \centering
    \includegraphics[width=0.8\linewidth]{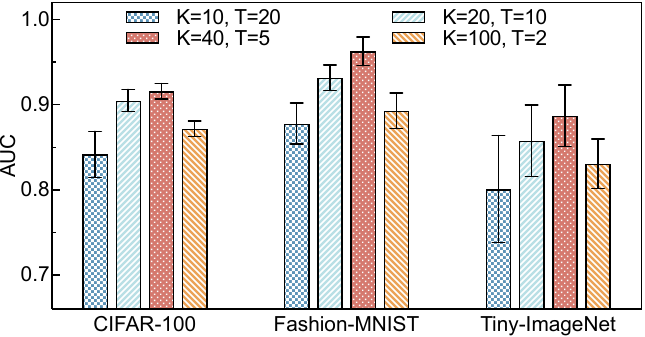}
    \caption{AUCs under different fingerprint numbers (K) and sample counts (T).}
    \label{fig:abl_KT_combo}
\end{figure}

We examine how the allocation between the number of fingerprints $K$ and the number of elements $T$ per fingerprint affects the verification performance when the total number of queries $K\times T$ is fixed at 200. 
As shown in Figure~\ref{fig:abl_KT_combo}, across three protected models (ResNet-18 on CIFAR-100, MobileNet-V2 on Fashion-MNIST, and ViT-B/16 on Tiny-ImageNet), the AUC varies notably across different $(K, T)$ combinations even though the total query budget remains constant. 
Specifically, the performance improves steadily as $K$ increases from 10 to 40 while $T$ decreases from 20 to 5, and then drops when $K$ further increases to 100 with $T=2$. 
For example, for ResNet-18 on CIFAR-100, the AUC rises from 0.842 at $(10,20)$ to 0.905 at $(20,10)$ and peaks at 0.916 for $(40,5)$, before declining to 0.872 at $(100,2)$. 
Similar trends are observed for MobileNet-V2 on Fashion-MNIST and ViT-B/16 on Tiny-ImageNet.
Overall, configurations with larger $K$ and smaller $T$ consistently outperform those with smaller $K$ and larger $T$ under the same total query budget. The configuration $(K=40, T=5)$ achieves the highest AUC across all settings under the fixed total query budget. 
This is because a larger $K$ allows the verification process to rely on a broader set of composite-sample fingerprints, which enhances the overall statistical reliability of the verification results. 

%%%%%%%%%%%%%%%%%%%%%%%%%%%%%%%%%%%%%%%%%%%%
\subsection{The impact of number of queries}
%%%%%%%%%%%%%%%%%%%%%%%%%%%%%%%%%%%%%%%%%%%%
\begin{figure}[htbp]
    \centering
    \includegraphics[width=0.8\linewidth]{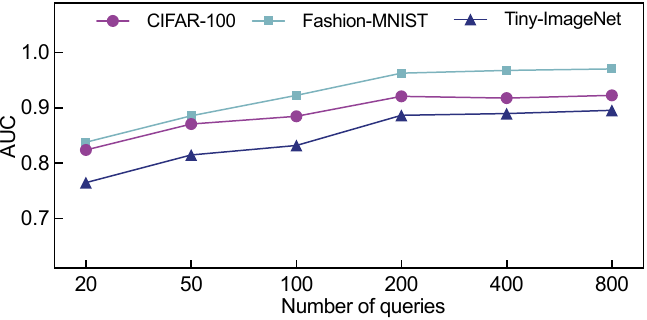}
   % \vspace{-0.1in}
    \caption{AUCs for different number of queries.}
    \label{fig:abl_query_num}
\end{figure}
We evaluate how the total number of queries used for verification affects the overall performance.
As shown in Figure~\ref{fig:abl_query_num}, the AUC increases steadily as the number of queries grows from 20 to 800 across all settings, demonstrating that using more queries improves verification reliability. 
However, once the number of queries reaches around 200, the AUC values become largely stable, and further increasing the query count yields only marginal gains. 
For instance, MobileNet-V2 on Fashion-MNIST, the AUC rises sharply from 0.838 at 20 queries to 0.963 at 200, while the improvement beyond 200 (up to 800 queries) is less than 0.8\%. 
This result suggests that a sufficient number of queries is essential for stable verification, but excessive queries provide diminishing returns. 
In practice, around 200 queries are sufficient to achieve near-optimal verification performance while maintaining computational efficiency.

\subsection{Impact of $\tau$}
\begin{figure}[htbp]
    \centering
    \includegraphics[width=0.8\linewidth]{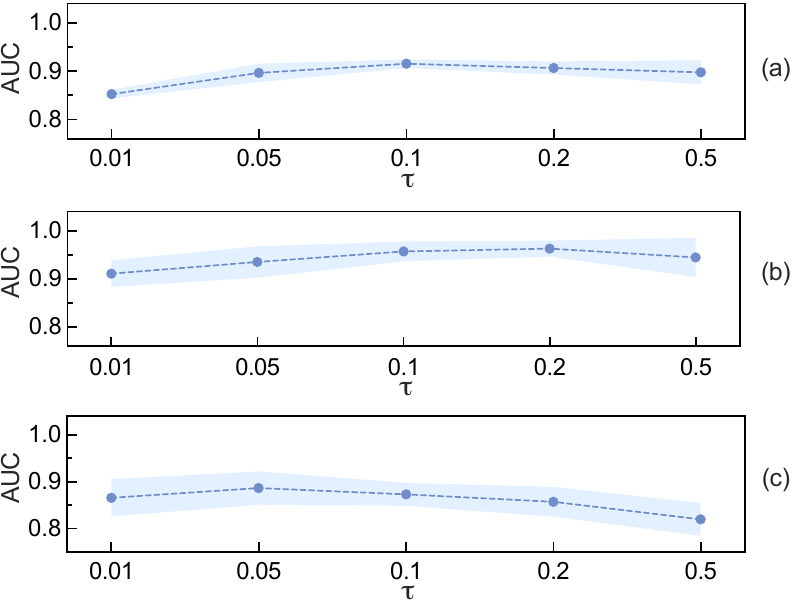}
   % \vspace{-0.1in}
    \caption{AUCs under different $\tau$ values for ResNet-18 on CIFAR-100 (a), MobileNet-V2 on Fashion-MNIST (b), and ViT-B/16 on Tiny-ImageNet (c).}
    \label{fig:abl_tau}
\end{figure}

We assess the impact of $\tau$ on verification performance. As shown in Figure~\ref{fig:abl_tau}, all three settings exhibit a similar trend: both small and large values of $\tau$ degrade AUC slightly, while a moderate one yields the best performance. This is because a small $\tau$ leads to fingerprints that are overly sensitive to decision boundary shifts, whereas a large one reduces the fingerprint's uniqueness. The optimal $\tau$ value depends on the total number of classes of each task: the larger the total number of classes, the smaller the value of $\tau$ should relatively be, and vice versa. The purpose of this scaling is to ensure that the fingerprints maintain a certain distance from multiple decision boundaries.
To improve the discriminative power, we need to optimize $\tau$ to find an intermediate value that achieves a balance between robustness and uniqueness.

\subsection{Impact of $\alpha$}
\begin{figure}[htbp]
    \centering
    \includegraphics[width=0.8\linewidth]{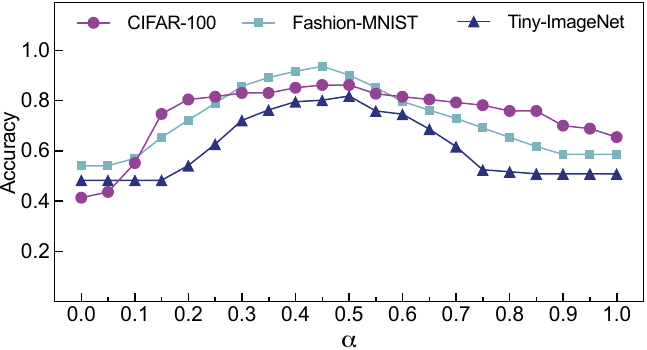}
   % \vspace{-0.1in}
    \caption{Accuracy for different $\alpha$.}
    \label{fig:abl_alpha}
\end{figure}

We study the impact of the verification threshold $\alpha$ on fingerprint verification performance across three different settings---ResNet-18 on CIFAR-100, MobileNet-V2 on Fashion-MNIST, and ViT-B/16 on Tiny-ImageNet.
The evaluation metric is the overall accuracy defined by
\begin{equation}
    \text{Accuracy}=\frac{TP + TN}{TP + FP + TN + FN},
\end{equation}
where TP, FP, TN, and FN denote true positive rate, false positive rate, true negative rate, and false negative rate, respectively.
As shown in Figure~\ref{fig:abl_alpha}, the verification accuracy varies as $\alpha$ changes.
For MobileNet-V2 on Fashion-MNIST, the accuracy increases steadily as the threshold moves from $\alpha=0$ toward the mid-range and reaches a peak of approximately $0.937$ when $\alpha$ is in the range 0.45\text{--}0.50, and then gradually declines as $\alpha$ becomes too large.
%A similar trend is observed on other two setting: 
Likewise, ResNet-18 and ViT-B/16 follow a similar trend: accuracy improves rapidly when $\alpha < 0.20$, remains near its optimum for $\alpha$ between $0.45$ and $0.50$, and deteriorates once $\alpha$ exceeds this range.

Overall, these results demonstrate that moderate threshold values (around $0.45\text{--}0.50$) provide the best verification performance.
Extremely low thresholds or extremely high thresholds lead to suboptimal accuracy, highlighting the importance of selecting an appropriate decision threshold for stable and reliable fingerprint verification.

\subsection{Impact of reference model set size}
\begin{figure}[htbp]
    \centering
    \includegraphics[width=0.8\linewidth]{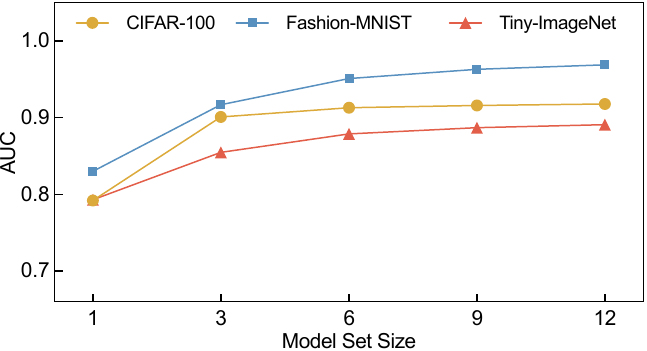}
   % \vspace{-0.1in}
    \caption{AUCs for different model set sizes.}
    \label{fig:abl_modelsize}
\end{figure}

We study the impact of reference model set size, i.e., the number of models in each of the reference model sets $\mathcal{V}_f$ and $\mathcal{I}_f$, on the verification performance. As shown in Figure~\ref{fig:abl_modelsize}, enlarging these sets consistently improves AUC across all settings---ResNet-18 on CIFAR-100, MobileNet-V2 on Fashion-MNIST, and ViT-B/16 on Tiny-ImageNet. 
For instance, for ResNet-18 on CIFAR-100, the AUC increases from 0.792 with one model per set to 0.918 with 12 models per set; for MobileNet-V2 on Fashion-MNIST, it increases from 0.830 to 0.969.
Particularly, the most substantial performance gains occur when increasing the set size from one to six, beyond which the performance is getting stable. These results indicate that our method can converge and achieve high stability of the verification performance with only moderately sized reference sets, underscoring its scalability and efficiency.

%%%%%%%%%%%%%%%%%%%%%%%%%%%%%%%%%%%%%%%%%%%%
\section{Time Overhead}
\label{sec:appendix_timeoverheads}
%%%%%%%%%%%%%%%%%%%%%%%%%%%%%%%%%%%%%%%%%%%%
IrisFP's time cost during the fingerprint generation process comes from two parts: (1) training the reference model set, and (2) generating the fingerprints.
We compare the time cost of using different model fingerprinting methods. For each method, we generate 200 fingerprints in total. It is noteworthy that IrisFP adopts 40 composite-sample fingerprints, each being composed of 5 samples, ensuring a fair assessment across different methods. Besides, we consider three protected models: a ResNet-18 trained on CIFAR-100, a MobileNet-V2 trained on Fashion-MNIST, and a ViT-B/16 trained on Tiny-ImageNet.

The experimental results are summarized in Table \ref{tab:time_overhead}. As shown by the table, IrisFP requires 55 $m$ for MobileNet-V2 on Fashion-MNIST, which is an acceptable cost among different methods. More importantly, even when switching to a substantially larger model ViT, the total cost only increases to 1 $h$ 32 $m$, demonstrating that the computational overhead of IrisFP scales modestly with the model size. This level of computational cost is practical and acceptable in real-world deployments. It is worth noting that for the time of performing ownership verification, we ignore the discussion. This is because querying a target model is nearly real-time, and the methods with the same total number of fingerprints incur same time cost.

\begin{table}[htbp]
% \vspace{-10pt}
    \centering
    \caption{Time overheads of generating fingerprints across different methods.}
    \label{tab:time_overhead}
    %\vspace{-0.1in}
    \resizebox{1.0\linewidth}{!}{ %
    \begin{tabular}{lccccc}
        \toprule
        \textbf{Method} & \textbf{IPGuard}          & \textbf{UAP}  & \textbf{ADV-TRA}  & \textbf{AKH} & \textbf{IrisFP}\\
        \midrule
        \textbf{ResNet-18}          & 59$s$         & 4$h$40$m$     & 32$m$        & 11$s$         & 1$h$2$m$ \\
        \textbf{MobileNet-V2}          & 41$s$         & 3$h$ 5$m$     & 24$m$        & 8$s$         & 55$m$ \\
        \textbf{ViT}                & 1$m$21$s$     & 7$h$30$m$     & 1$h$11$m$    & 1$m$6$s$      & 1$h$32$m$\\
        \bottomrule
    \end{tabular}
    }
% \vspace{-10pt}
\end{table}

%%%%%%%%%%%%%%%%%%%%%%%%%%%%%%%%%%%%%%%%%%%%
\section{Stealthiness}
\label{sec:appendix_stealthiness}
\begin{figure}[htbp]
    \centering
    \includegraphics[width=0.8\linewidth]{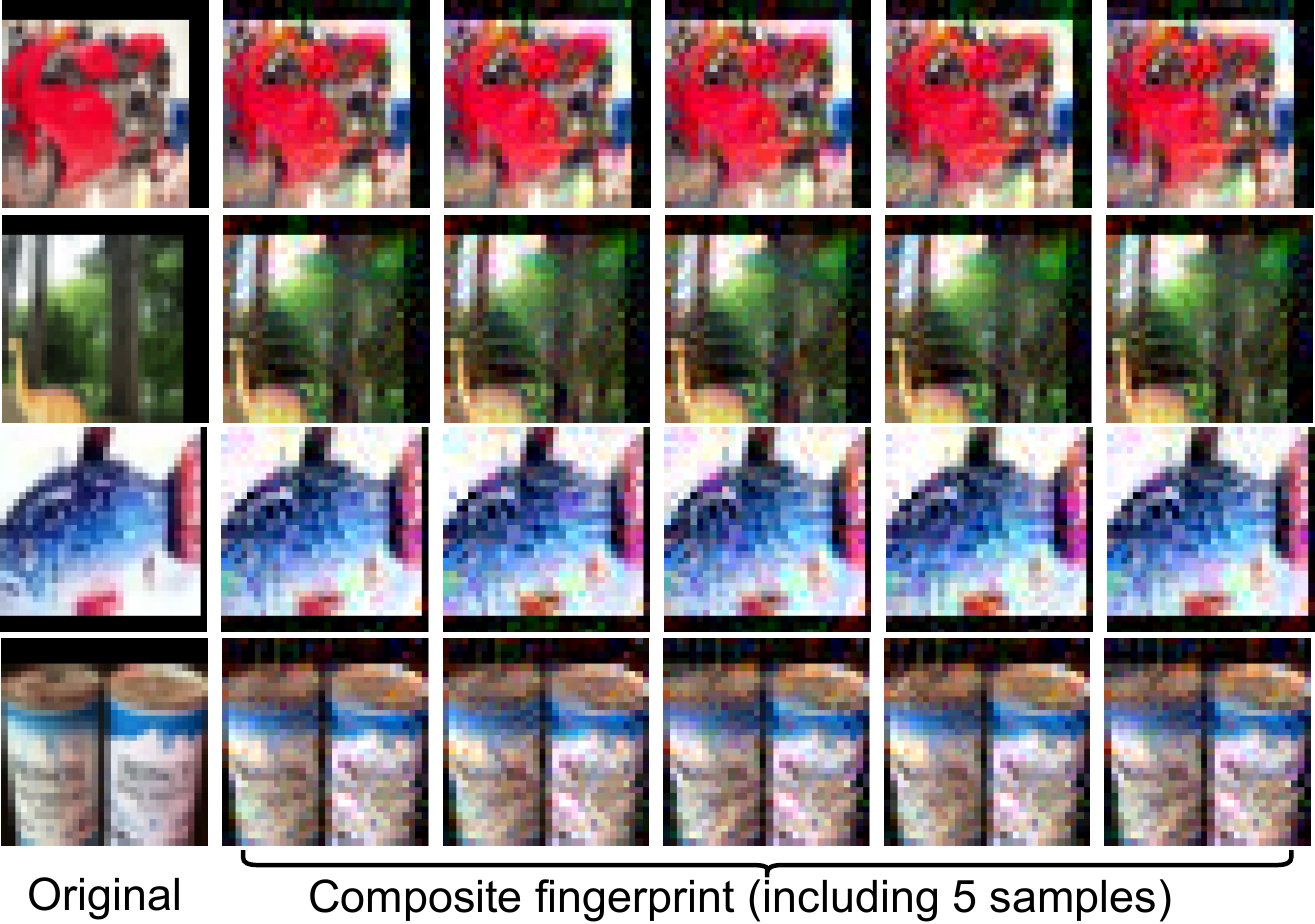}
    \caption{Qualitative evaluation of stealthiness. Each row shows an original sample and its corresponding composite fingerprint consists of five samples.}
    \label{fig:stealthiness}
\end{figure}

%%%%%%%%%%%%%%%%%%%%%%%%%%%%%%%%%%%%%%%%%%%%
Prior fingerprinting studies rarely report ``stealthiness/imperceptibility'', as enforcing extremely small visual perturbations often weakens the discriminative strength of the fingerprints. Nevertheless, we add a qualitative evaluation of stealthiness in Fig.~\ref{fig:stealthiness}. 
As shown in the figure, the generated fingerprints remain visually similar to the original inputs, with only subtle pixel-level differences.

%%%%%%%%%%%%%%%%%%%%%%%%%%%%%%%%%%%%%%%%%%%%
\section{Task Transferability}
\label{sec:appendix_tasktransfer}
%%%%%%%%%%%%%%%%%%%%%%%%%%%%%%%%%%%%%%%%%%%%
IrisFP targets the standard ownership-verification setting considered in prior work, where the target model is deployed for the same task and shares the same label set as the protected model, even after model-modification attacks. If the task changes and the label space differs, verification can still be conducted using the overlapping subset of labels, since our fingerprint set spans multiple labels. However, the verification performance is expected to degrade as the degree of label overlap decreases.

%%%%%%%%%%%%%%%%%%%%%%%%%%%%%%%%%%%%%%%%%%%%%%%
\section{Experimental Setting Details}
\label{sec:appendix_expsetting}
%%%%%%%%%%%%%%%%%%%%%%%%%%%%%%%%%%%%%%%%%%%%%%%

\subsection{Protected Model}
We consider three model architectures for the protected model: ResNet-18, MobileNet-V2, and ViT-B/16. For each of the five benchmark datasets (CIFAR-10, CIFAR-100, Fashion-MNIST, MNIST, and Tiny-ImageNet), we independently train two protected models: one using ResNet-18 and the other using MobileNet-V2. Besides, we also produce a protected model with the ViT-B/16 architecture, a larger and more complex architecture, by finetuning it on Tiny-ImageNet. All ResNet-18 and MobileNet-V2 models follow standard training configurations---SGD with momentum 0.9, weight decay of $5\times10^{-4}$, an initial learning rate of 0.1, cosine-annealing scheduling, Xavier initialization, and a batch size of 128 for 600 epochs. For ViT-B/16, which is generally more suitable for higher-resolution data, we initialize the model from pretrained weights and finetune it on Tiny-ImageNet under the same settings, except that we use a smaller initial learning rate of 0.01 and train it for 50 epochs.

\subsection{Reference Model Sets}
Two reference model sets, including the pirated model set and the independently-trained set, serve as the foundation for experimentally assessing the quality of fingerprints during the refinement process. 
For each protected model, its corresponding pirated model set is constructed by modifying the protected model through three model modification techniques, chosen from six available attack types (FT, PR, KD, AT, PFT, and NFT). These modifications invalidate the original fingerprints, resulting in evading ownership verification.
We adopt only three of them to balance computational cost and to emulate realistic conditions where reference models may not encompass all possible removal attempts. In our experiment, we select fine-tuning (FT), knowledge distillation (KD), and adversarial training (AT).
Specifically:
\begin{itemize}
    \item Fine-tuning (FT): Training the protected model for a specified number of epochs.
    \item Knowledge Distillation (KD): Training a student model---either with the same or a different architecture---under the supervision of the protected (teacher) model by minimizing the KL divergence between their soft outputs, with temperature set to 1.
    \item Adversarial Training (AT): Crafting adversarial examples via a PGD attack (using Foolbox’s ``LinfPGD'' tool), concatenating them with clean inputs, and then training the model on the mixed dataset.
\end{itemize}
For each attack type, we produce three pirated variants instantiated with 3 different random seeds to ensure model, leading to 9 pirated variants in the pirated model set.
On the other hand, the easy way for constructing the independent model set is to use public models from an open-source platform, e.g., Hugging Face. In our experiments,  the model is trained from scratch with different seeds and hyperparameters from training the protected models, following standard training configurations---SGD with momentum 0.9, weight decay of $5\times10^{-4}$, an initial learning rate of 0.1, cosine-annealing scheduling, Xavier initialization, and a batch size of 128 for 600 epochs.
Specifically, our independent models are constructed by using three diverse architectures available in the torchvision library---ResNet-18, MobileNet-V2, and DenseNet-121---each type instantiated with 3 different random seeds to create 3 independently-trained models, resulting in a total of 9 models in the independent model set. To guarantee independence, all ResNet-18 and MobileNet-V2 instances are initialized with seeds distinct from that of the protected model.

\subsection{Testing Model Set}
The testing model set consists of pirated and independently-trained models, which are independently trained and have no overlap with the models in the reference model sets. 

The pirated models in the testing set are constructed using six representative attacks:
\begin{itemize}
    \item Fine-tuning (FT): Training the protected model for a specified number of epochs.
    \item Pruning (PR): Applying unstructured global pruning with sparsity levels ranging from 10\% to 90\% in increments of 10\%, without retraining.
    \item Prune-then-tune (PFT): Applying pruning at sparsity levels of 30\%, 60\%, and 90\%, followed by fine-tuning.
    \item Noise-regularized Fine-tuning (NFT): Perturbing each trainable parameter tensor with Gaussian noise scaled by a scaled magnitude of standard deviation of that layer, i.e., $param+= \alpha \cdot std(param) \cdot \mathcal{N}(0,1)$, followed by fine-tuning.
    \item Knowledge Distillation (KD): Training a student model---of the same or a different architecture---using the KL divergence between its outputs and those of the protected model (temperature = 1).
    \item Adversarial Training (AT): Crafting adversarial examples via a PGD attack (using Foolbox’s ``LinfPGD'' tool), concatenating them with clean inputs, and then training the model on the mixed dataset.
\end{itemize}
For each attack type, we generate 20 variants with different random seeds, resulting in a total 120 pirated models.

The independently-trained models in the testing set are built from scratch using six architectures: ResNet-18, ResNet-50, MobileNet-V2, MobileNet-V3 Large, EfficientNet-B2, and DenseNet-121. For each architecture, 20 models are trained with distinct seeds, without any access to the protected models or the models in the reference model set, yielding 120 independently-trained models.

\subsection{Hyperparameters}

We use $K=40$ composite-sample fingerprints, each consisting of $T=5$ samples for all tasks.
The bias parameter $\tau$ is 0.2 for CIFAR-10, Fashion-MNIST, and MNIST, 0.1 for CIFAR-100, and 0.05 for Tiny-ImageNet.
We set the regularization coefficients $\lambda_1$ and $\lambda_2$ to 0.05.
The training process consists of 800 iterations in Phase I and 200 iterations in Phase II.

\subsection{Computing Infrastructure}
All experiments were run on a Linux server equipped with an NVIDIA A100 GPU and 200 GB of system memory. GPU acceleration was provided by CUDA 12.1, and all models and training pipelines were implemented in PyTorch 2.5.1.

%%%%%%%%%%%%%%%%%%%%%%%%%%%%%%%%%%%%%%%%%%%%
\section{Related Work}
\label{sec:appendix_relatedwork}
%%%%%%%%%%%%%%%%%%%%%%%%%%%%%%%%%%%%%%%%%%%%
% \subsection{Model Fingerprinting}
Model fingerprinting has emerged as a promising non-intrusive technique for protecting the intellectual property (IP) of deep neural networks (DNNs), which are highly susceptible to unauthorized use \cite{AAFA20, DNNF21, IPIP21, AYSM22, GAFF24, CNNLT22, BUNNF25, EMIPPR25}. 
By exploiting the distinctive behavioral patterns of a protected model, fingerprinting derives verifiable ownership evidence\cite{TMFA21,DMFA24}.
% \cite{CEHL21,IPIP21,MFDN22,NGNF23,GGFG23,MCGF24,UWSDW24}. 

Researchers have developed various model fingerprinting methods. For example, several fingerprinting methods assess model similarity through internal representations or model parameters instead of adversarial queries.
Maho \etal~\cite{FCBI23} propose a greedy scheme that generates fingerprints for a model by analyzing the statistical similarity of intermediate activations from benign inputs using Shannon’s information theory.
Guan \etal~\cite{AYSM22} identify suspicious models by selecting inputs that yield inconsistent predictions across two sets of reference models and examining their pairwise relationships.
Other methods treat the model parameters themselves as intrinsic fingerprints. For instance,
Jia \etal~\cite{jia2021zest} train linear proxy models on a reference dataset and compare their learned weights via cosine similarity, while Zheng \etal~\cite{zheng2022dnn} project front-layer weights into a random subspace associated with the model owner’s identity to achieve non-repudiable ownership verification.

In the recent years, researchers have gained great attention to the adversarial-example-based fingerprinting and have developed many methods for it. This type of fingerprinting leverages adversarial examples to craft input–output pairs that exhibit model-specific behavioral patterns. For instance, Cao \etal~\cite{IPIP21} reveal that a model’s decision boundary reflects its unique identity and generates data points (adversarial samples) near a decision boundary to serve as distinctive fingerprints. Lukas \etal~\cite{DNNF21} optimizes adversarial examples through multi-model training to maximize their transferability to pirated models. However, such transferability often leads to false positives.
%as independent models may share similar adversarial examples.
To address this, Peng \etal~\cite{FDNN22} employ Universal Adversarial Perturbations to modify clean inputs into fingerprint samples, and then use a learned encoder to map model logits into embedding vectors for similarity-based ownership verification.
Yang \etal~\cite{NGNF23} train a GAN to synthesize natural fingerprint samples by optimizing inputs that yield divergent predictions between pirated and independent models in decision-difference regions, which are then used as black-box queries for verification.
Liu \etal~\cite{MCGF24} leverage a conditional GAN with margin loss to generate samples positioned at controlled distances from the classification boundary, enabling robust and distinctive fingerprinting without relying on surrogate models.
Instead of isolated adversarial samples, Xu \etal~\cite{UWSD24} construct adversarial trajectories by iteratively perturbing clean inputs to traverse decision-boundary regions, capturing richer and more comprehensive model-specific behaviors.
Godinot \etal~\cite{QURD25} introduce a decomposition-based analysis framework that separates fingerprinting into query construction, representation, and detection, and show that one effective instantiation constructs fingerprints using adversarial examples originating from naturally misclassified samples.

Despite these advancements, a fundamental challenge remains: designing fingerprints that simultaneously achieve uniqueness—distinguishing the protected model from independently trained ones—and robustness—maintaining validity under model modifications or removal attacks.
Different from conventional approaches that typically construct each fingerprint near a single decision boundary, we investigate the region near the intersection of multiple decision boundaries and propose a composite-sample fingerprinting framework that jointly enhances both uniqueness and robustness.

%% file: main.bib
@String(CVPR= {IEEE Conf. Comput. Vis. Pattern Recog.})

@String(ICASSP=	{ICASSP})

@String(ICLR = {Int. Conf. Learn. Represent.})

@String(AAAI = {AAAI})

@String(CVPR  = {CVPR})

@String(ICLR  = {ICLR})

@inproceedings{CRAT21,
  title={{Copy, Right? A} Testing Framework for Copyright Protection of Deep Learning Models},
  author={Jialuo Chen and Jingyi Wang and Tinglan Peng and Youcheng Sun and Peng Cheng and Shouling Ji and Xingjun Ma and Bo Li and Dawn Xiaodong Song},
  booktitle={Proceedings of IEEE Symposium on Security and Privacy (S\&P'22)},
  year={2022},
  pages={824-841}
}

@inproceedings{IPIP21,
  title={{IPGuard}: Protecting Intellectual Property of Deep Neural Networks via Fingerprinting the Classification Boundary},
  author={Xiaoyu Cao and Jinyuan Jia and Neil Zhenqiang Gong},
  booktitle={Proceedings of the 2021 ACM Asia Conference on Computer and Communications Security (AsiaCCS'21)},
  year={2021}
}

@article{AAFA20,
  title={{AFA}: Adversarial fingerprinting authentication for deep neural networks},
  author={Zhao, Jingjing and Hu, Qingyue and Liu, Gaoyang and Ma, Xiaoqiang and Chen, Fei and Hassan, Mohammad Mehedi},
  journal={Computer Communications},
  volume={150},
  pages={488--497},
  year={2020},
  publisher={Elsevier}
}

@inproceedings{DNNF21,
title={Deep Neural Network Fingerprinting by Conferrable Adversarial Examples},
author={Nils Lukas and Yuxuan Zhang and Florian Kerschbaum},
booktitle={International Conference on Learning Representations (ICLR'21)},
year={2021}
}

@inproceedings{TMFA21,
author = {Yufei Chen and Chao Shen and Cong Wang and Yang Zhang},
title = {Teacher Model Fingerprinting Attacks Against Transfer Learning},
booktitle = {Proceedings of 31st USENIX Security Symposium (USESec'22)},
year = {2022},
address = {Boston, MA},
pages = {3593--3610}
}

@inproceedings{FDNN22,
  title={Fingerprinting Deep Neural Networks Globally via Universal Adversarial Perturbations},
  author={Zirui Peng and Shaofeng Li and Guoxing Chen and Cheng Zhang and Haojin Zhu and Minhui Xue},
  booktitle={Proceedings of 2022 IEEE/CVF Conference on Computer Vision and Pattern Recognition (CVPR'22)},
  year={2022},
  pages={13420-13429}
}

@inproceedings{AYSM22,
title={Are You Stealing My Model? Sample Correlation for Fingerprinting Deep Neural Networks},
author={Jiyang Guan and Jian Liang and Ran He},
booktitle={Proceedings of Advances in Neural Information Processing Systems (NeurIPS'22)},
editor={Alice H. Oh and Alekh Agarwal and Danielle Belgrave and Kyunghyun Cho},
year={2022}
}

@article{NGNF23,
  title={{NaturalFinger}: Generating Natural Fingerprint with Generative Adversarial Networks},
  author={Kan Yang and Kunhao Lai},
  journal={arXiv preprint arXiv:2305.17868},
  year={2023}
}

@article{DMFA24,
  title={Dual-verification-based model fingerprints against ambiguity attacks},
  author={Zhao, Boyao and Chen, Haozhe and Zhang, Jie and Zhang, Weiming and Yu, Nenghai},
  journal={Cybersecurity},
  volume={7},
  number={1},
  pages={78},
  year={2024},
  publisher={Springer}
}

@inproceedings{GAFF24,
  title={{GNNFingers}: A Fingerprinting Framework for Verifying Ownerships of Graph Neural Networks},
  author={Xiaoyu You and Youhe Jiang and Jianwei Xu and Mi Zhang and Min Yang},
  booktitle={Proceedings of the ACM on Web Conference (WWW'24)},
  year={2024}
}

@inproceedings{MCGF24,
  title={{MarginFinger}: Controlling Generated Fingerprint Distance to Classification boundary Using Conditional GANs},
  author={Weixing Liu and Shenghua Zhong},
  booktitle={Proceedings of the 2024 International Conference on Multimedia Retrieval (ICMR'24)},
  year={2024}
}

@inproceedings{UWSD24,
  title={United We Stand, Divided We Fall: Fingerprinting Deep Neural Networks via Adversarial Trajectories},
  author={Tianlong Xu and Chen Wang and Gaoyang Liu and Yang Yang and Kai Peng and Wei Liu},
  booktitle={Proceedings of Neural Information Processing Systems  (NeurIPS'24)},
  year={2024}
}

@inproceedings{QURD25,
  title={Queries, Representation \& Detection: The Next 100 Model Fingerprinting Schemes},
  author={Godinot, Augustin and Le Merrer, Erwan and Penzo, Camilla and Ta{\"\i}ani, Fran{\c{c}}ois and Tr{\'e}dan, Gilles},
  booktitle={Proceedings of the AAAI Conference on Artificial Intelligence (AAAI'25)},
  volume={39},
  number={16},
  pages={16817--16825},
  year={2025}
}

@inproceedings{PFEC17,
  author={Hao Li and Asim Kadav and Igor Durdanovic and Hanan Samet and Hans Peter Graf},
  title={Pruning Filters for Efficient ConvNets},
  year={2017},
  booktitle={ICLR'17}
}

@article{CSTL20,
  title={A comprehensive survey on transfer learning},
  author={Zhuang, Fuzhen and Qi, Zhiyuan and Duan, Keyu and Xi, Dongbo and Zhu, Yongchun and Zhu, Hengshu and Xiong, Hui and He, Qing},
  journal={Proceedings of the IEEE},
  volume={109},
  number={1},
  pages={43--76},
  year={2020},
  publisher={IEEE}
}

@article{CIFAR10,
  title={The {CIFAR-10} dataset},
  author={Krizhevsky, Alex and Nair, Vinod and Hinton, Geoffrey and others},
  journal={online: http://www. cs. toronto. edu/kriz/cifar. html},
  volume={55},
  number={5},
  pages={2},
  year={2014}
}

@article{CIFAR100,
  title={The {CIFAR-100} dataset},
  author={Krizhevsky, Alex and Nair, Vinod and Hinton, Geoffrey and others},
  journal={online: http://www. cs. toronto. edu/kriz/cifar. html},
  volume={55},
  number={5},
  pages={2},
  year={2014}
}

@article{MNIST,
  title={{MNIST} handwritten digit database},
  author={LeCun, Yann and Cortes, Corinna and Burges, CJ},
  journal={ATT Labs [Online]. Available: http://yann.lecun.com/exdb/mnist},
  volume={2},
  year={2010}
}

@article{FMNIST,
  title={Fashion-MNIST: a Novel Image Dataset for Benchmarking Machine Learning Algorithms},
  author={Han Xiao and Kashif Rasul and Roland Vollgraf},
  journal={ArXiv},
  year={2017},
  volume={abs/1708.07747},
  url={https://api.semanticscholar.org/CorpusID:702279}
}

@inproceedings{TinyImageNet,
  title={Tiny ImageNet Visual Recognition Challenge},
  author={Ya Le and Xuan S. Yang},
  year={2015},
  url={https://api.semanticscholar.org/CorpusID:16664790}
}

@inproceedings{HAAAH20,
  title={High accuracy and high fidelity extraction of neural networks},
  author={Jagielski, Matthew and Carlini, Nicholas and Berthelot, David and Kurakin, Alex and Papernot, Nicolas},
  booktitle={29th USENIX security symposium (USESec'20)},
  pages={1345--1362},
  year={2020}
}

@inproceedings{CNNLT22,
  title={Can neural nets learn the same model twice? investigating reproducibility and double descent from the decision boundary perspective},
  author={Somepalli, Gowthami and Fowl, Liam and Bansal, Arpit and Yeh-Chiang, Ping and Dar, Yehuda and Baraniuk, Richard and Goldblum, Micah and Goldstein, Tom},
  booktitle={Proceedings of the IEEE/CVF Conference on Computer Vision and Pattern Recognition (CVPR'22)},
  pages={13699--13708},
  year={2022}
}

@inproceedings{UWSDW24,
title={United We Stand, Divided We Fall: Fingerprinting Deep Neural Networks via Adversarial Trajectories},
author={Tianlong Xu and Chen Wang and Gaoyang Liu and Yang Yang and Kai Peng and Wei Liu},
booktitle={The Thirty-eighth Annual Conference on Neural Information Processing Systems (NeurIPS'24)},
year={2024},
}

@inproceedings{IPAGM24,
  title={{IPRemover}: A generative model inversion attack against deep neural network fingerprinting and watermarking},
  author={Zong, Wei and Chow, Yang-Wai and Susilo, Willy and Baek, Joonsang and Kim, Jongkil and Camtepe, Seyit},
  booktitle={Proceedings of the AAAI Conference on Artificial Intelligence (AAAI'24)},
  volume={38},
  number={7},
  pages={7837--7845},
  year={2024}
}

@inproceedings{DCCDN15,
  title={Deep compression: Compressing deep neural networks with pruning, trained quantization and huffman coding},
  author={Han, Song and Mao, Huizi and Dally, William J},
  booktitle={ICLR'16},
  year={2016}
}

@inproceedings{TFVVI23,
  title={Task-specific fine-tuning via variational information bottleneck for weakly-supervised pathology whole slide image classification},
  author={Li, Honglin and Zhu, Chenglu and Zhang, Yunlong and Sun, Yuxuan and Shui, Zhongyi and Kuang, Wenwei and Zheng, Sunyi and Yang, Lin},
  booktitle={Proceedings of the IEEE/CVF Conference on Computer Vision and Pattern Recognition (CVPR'23)},
  pages={7454--7463},
  year={2023}
}

@article{BMRVI24,
  title={Boosting model resilience via implicit adversarial data augmentation},
  author={Zhou, Xiaoling and Ye, Wei and Lee, Zhemg and Xie, Rui and Zhang, Shikun},
  journal={arXiv preprint arXiv:2404.16307},
  year={2024}
}

@article{ATFF19,
  title={Adversarial training for free!},
  author={Shafahi, Ali and Najibi, Mahyar and Ghiasi, Mohammad Amin and Xu, Zheng and Dickerson, John and Studer, Christoph and Davis, Larry S and Taylor, Gavin and Goldstein, Tom},
  journal={Advances in Neural Information Processing Systems (NeurIPS'19)},
  volume={32},
  year={2019}
}

@inproceedings{OFTA24,
  title={Out of thin air: Exploring data-free adversarial robustness distillation},
  author={Wang, Yuzheng and Chen, Zhaoyu and Yang, Dingkang and Guo, Pinxue and Jiang, Kaixun and Zhang, Wenqiang and Qi, Lizhe},
  booktitle={Proceedings of the AAAI Conference on Artificial Intelligence (AAAI'24)},
  volume={38},
  number={6},
  pages={5776--5784},
  year={2024}
}

@inproceedings{MEAR24,
  title={Model extraction attacks revisited},
  author={Liang, Jiacheng and Pang, Ren and Li, Changjiang and Wang, Ting},
  booktitle={Proceedings of the 19th ACM Asia Conference on Computer and Communications Security (AsiaCCS'24)},
  pages={1231--1245},
  year={2024}
}

@inproceedings{IBSF24,
  title={Intersecting-boundary-sensitive fingerprinting for tampering detection of {DNN} models},
  author={Xiaofan Bai and He, Chaoxiang and Ma, Xiaojing and Zhu, Bin Benjamin and Jin, Hai},
  booktitle={Proceedings of Forty-first International Conference on Machine Learning (ICML'24)},
  year={2024}
}

@inproceedings{SDBF25,
  title={{SDBF}: Steep-Decision-Boundary Fingerprinting for Hard-Label Tampering Detection of {DNN} Models},
  author={Bai, Xiaofan and Li, Shixin and Ma, Xiaojing and Zhu, Bin Benjamin and Zhang, Dongmei and Yu, Linchen},
  booktitle={Proceedings of the Computer Vision and Pattern Recognition Conference (CVPR'25)},
  pages={29278--29287},
  year={2025}
}

@inproceedings{BUNNF25,
  title={Boosting the Uniqueness of Neural Networks Fingerprints with Informative Triggers},
  author={Zhang, Zhuomeng and Li, Fangqi and Wang, Hanyi and Wang, Shi-Lin},
  booktitle={Proceedings of The Thirty-ninth Annual Conference on Neural Information Processing Systems (NeurIPS'25)},
  year={2025}
}

@article{EMIPPR25,
  title={Enhancing Model Intellectual Property Protection with Robustness Fingerprint Technology},
  author={Yan, Anli and Ren, Huali and Mo, Kanghua and Zhang, Zhenxin and Wang, Shaowei and Li, Jin},
  journal={IEEE Transactions on Information Forensics and Security (TIFS)},
  year={2025},
  publisher={IEEE}
}

@article{FITP25,
  title={Fit-print: Towards false-claim-resistant model ownership verification via targeted fingerprint},
  author={Shao, Shuo and Zhu, Haozhe and Li, Yiming and Yao, Hongwei and Zhang, Tianwei and Qin, Zhan},
  journal={arXiv preprint arXiv:2501.15509},
  year={2025}
}

@inproceedings{RFRF2025,
  title={Rethinking the Fragility and Robustness of Fingerprints of Deep Neural Networks},
  author={Li, Fangqi and Wang, Shilin and Yang, Lei},
  booktitle={Proceedings of IEEE International Conference on Acoustics, Speech and Signal Processing (ICASSP'25)},
  pages={1--5},
  year={2025},
  organization={IEEE}
}

@article{lin2025adversarial,
  title={Adversarial Example Based Fingerprinting for Robust Copyright Protection in Split Learning},
  author={Lin, Zhangting and Xue, Mingfu and Chen, Kewei and Liu, Wenmao and Gao, Xiang and Zhang, Leo Yu and Wang, Jian and Zhang, Yushu},
  journal={arXiv preprint arXiv:2503.04825},
  year={2025}
}

@ARTICLE{FCBI23,
  author={Maho, Thibault and Furon, Teddy and Merrer, Erwan Le},
  journal={IEEE Transactions on Information Forensics and Security (TIFS)}, 
  title={Fingerprinting Classifiers With Benign Inputs}, 
  year={2023},
  volume={18},
  number={},
  pages={5459-5472}}

@inproceedings{jia2021zest,
  title={A zest of lime: Towards architecture-independent model distances},
  author={Jia, Hengrui and Chen, Hongyu and Guan, Jonas and Shamsabadi, Ali Shahin and Papernot, Nicolas},
  booktitle={International Conference on Learning Representations (ICLR'21)},
  year={2021}
}

@article{zheng2022dnn,
  title={A DNN fingerprint for non-repudiable model ownership identification and piracy detection},
  author={Zheng, Yue and Wang, Si and Chang, Chip-Hong},
  journal={IEEE Transactions on Information Forensics and Security (TIFS)},
  volume={17},
  pages={2977--2989},
  year={2022},
  publisher={IEEE}
}

@inproceedings{
yang2026anafp,
title={Fingerprinting Deep Neural Networks for Ownership Protection: An Analytical Approach},
author={Guang Yang and Ziye Geng and Yihang Chen and Changqing Luo},
booktitle={The Fourteenth International Conference on Learning Representations (ICLR'26)},
year={2026},
url={https://openreview.net/forum?id=sg3UNWKVFt}
}

@inproceedings{
yang2026liteguard,
title={LiteGuard: Efficient Task-Agnostic Model Fingerprinting with Enhanced Generalization},
author={Guang Yang and Ziye Geng and Yihang Chen and Changqing Luo},
booktitle={The Fourteenth International Conference on Learning Representations (ICLR'26)},
year={2026},
url={https://openreview.net/forum?id=TFC25ZT9nI}
}
